\documentstyle[12pt]{article}
\def\half{{ 1\over 2}}
\def\Half{{1/2}}
\def\n{\noindent}
\def\vp{\varphi}
\def\al{\alpha}
\def\b{\beta}
\def\k{\kappa}
\def\L{\Lambda}
\def\sr{\sqrt}
\def\en{\eqno}
\def\th{\theta}
\def\G{\Gamma}
\def\ve{\varepsilon}
\def\eps{\varepsilon}
\def\e{\epsilon}
\def\l{\lambda}
\def\d{\delta}
\def\z{\zeta}
\def\pt{\partial}
\newcommand{\arctg}{{\rm arctg}}
\newcommand{\tg}{{\rm tg}}
\newcommand{\Le}{\frac{\lambda}{4e^2}}
\newcommand{\Tr}{{\bf Tr}}
\newcommand{\rot}{{\rm rot}}
\newcommand{\vu}{{\bf u}}
\newcommand{\vn}{{\bf n}}
\newcommand{\vphi}{\varphi}

\newcommand{\GG}{{\bf G}}
\newcommand{\WW}{{\bf W}}
\newcommand{\EE}{{\bf E}}
\newcommand{\UU}{{\bf U}}
\newcommand{\V}{{\bf V}}
\newcommand{\RR}{{\bf R}}
\newcommand{\BB}{{\bf B}}
\newcommand{\DD}{{\bf D}}
\newcommand{\I}{{\bf I}}
\newcommand{\JJ}{{\bf JJ}}
\newcommand{\R}{{\bf R}}
\newcommand{\M}{{\bf M}}
\newcommand{\A}{{\bf A}}
\newcommand{\PP}{{\bf P}}
\newcommand{\QQ}{{\bf Q}}
\newcommand{\LL}{{\bf L}}
\newcommand{\xx}{{\bf x}}
\newcommand{\beq}[1]{\begin{equation} \label{#1}}
\newcommand{\eeq}{\end{equation}}
\newcommand{\bqa}[1]{\begin{eqnarray*} \label{#1}}
\newcommand{\eqa}{\end{eqnarray*}}
\newcommand{\beqa}[1]{\begin{eqnarray} \label{#1}}
\newcommand{\eeqa}{\end{eqnarray}}
\newcommand{\dx}[3]{\frac{\partial^{#1}{#2}}{\partial{x^#3}^{#1}}}
\newcommand{\dt}[2]{\frac{\partial^{#1}{#2}}{\partial{t}^{#1}}}
\newcommand{\rf}[1]{(\ref{#1})}
\newcommand{\Df}[1]{{{\bf D}^{#1}_\xi}}
\newcommand{\cz}{{\bar \zeta}}
\newcommand{\Dp}[1]{\frac{\partial}{\partial{#1}}}
\begin{document}
%\maketitle

%%%%%%%%%%%%%%%%%%%%%

%\heads{Chervon, Zhuravlev and Shchigolev}
%{New exact solutions in standard inflationary models.}

%\bls{1.01}

%%%%%%%%%%%%%%%%%%%%%%%%%%%%%%%%%%%%%%%%%%%%%%%%%%%%%%%%%%%%%%%%%%%%
%\begin{document}
%\thispagestyle{empty}
%\twocolumn[
%\noi \unitlength=1mm
\begin{picture}(174,8)
   \put(31,8){\shortstack[c]
       {RUSSIAN GRAVITATIONAL SOCIETY                \\
       ULYANOVSK STATE UNIVERSITY        }       }
\end{picture}
\begin{flushright}
                                         RGS-USU-97/02 \\
                                         gr-qc/9706031    \\
%	{\it Grav. and Cosmol.} {\bf 2}, 3\foom 1, 221-226 (1996)
\end{flushright}
%\bigskip
\begin{center}
{\large{NEW EXACT SOLUTIONS IN STANDARD INFLATIONARY MODELS}}
\end{center}
\begin{center}
{\it {S.V.Chervon, V.M.Zhuravlev, V.K.Shchigolev}}
\end{center}
%\thanks{e-mail: chervon@themp.univ.simbirsk.su}}
\date{}

\begin{center}
Department of Theoretical and Mathematical Physics,\
Ulyanovsk State University, \\
42, Leo Tolstoy Str., 432700 Ulyanovsk, Russia \
Tel/FAX:007-8422-344338;\\
E-mail: chervon@themp.univ.simbirsk.su\
\end{center}

\abstract{The exact solutions in the standard inflationary model based on the
self-interacting scalar field minimally coupled to gravity are considered.
The shape's freedom of the self-interacting potential $V(\phi)$ is
postulated to obtain a new set of the exact solutions in the framework of
Friedmann-Robertson-Walker Universes. The general solution was found in the
case of power law inflation. We obtained new solutions and compared them with
obtained ones earlir for the exponential type inflation.
}
\vspace{1.3cm}
\begin{center}
\end{center}
\vspace{1.3cm}
\begin{center}
\end{center}
\vfill\eject
\section{Introduction}
Cosmological inflationary models can be understood as the self-consistent
system of Einstein's and nonlinear (scalar, chiral, gauge) fields'
equations. Inflationary scenarios are usually based on this type system of
equations and contain an analysis of physical reasons which are the sources
of an extremely quick expansion (inflation) of the Universe. As a rule
the inflationary scenarios are connected with the phase transitions in
the framework of quantum theory of scalar field of a finite temperature
as well as with spontaneous violation and restoration of the gauge symmetry
and another physical phenomenon.

Various type of approximations and numerical methods have been actively
used in analysis of inflationary models and brought in the early 80-th
the bright success in
the explanation of the long standing problems of the Big Bang model such
as a horizon, monopolies, large scale structure formation \cite{guth81}-
\cite{albste82}.
Nevertheless now we find out the necessity to make sure in used approximations
to match the inflationary model to large scale structure of the Universe.
The best way to compare numerical or approximated results with real ones
based on the exploitation of the exact solutions.

In the work by A.Guth \cite{guth81} the inflationary solution has been
obtained with the potential $V(\phi)=const$.
The exponential expansion of the Universe as an example of the exact solution
has been found by G.Ivanov \cite{ivanov80} for
a nonlinear scalar field with the potential
$V(\phi)=\frac{\mu}{2}\phi^2-\frac{\l}{4}\phi^4 $
in the framework of spatially-flat Friedmann-Robertson-Walker (FRW) spaces
and has been interpretated as the Universe started from a quasivacuum state
of matter.
Exact solutions of the power law type inflation have been
obtained for Liouville non-linearity $V(\phi)=me^{-\l\phi}$ \cite{bbl86}
(see also \cite{barsai93} and references quoted therein).
The Liouville-type and some other non-linearities have been investigated
and some exact and asymptotic solutions are presented in \cite{muslimov90}.
Classical de Sitter type solutions was obtained in \cite{ellmad91}.
New classes of exact solutions have been found in \cite{barrow94} by
taking the scalar field as the function of time $\phi = \phi (t)$ and then
determining the evolution of the expansion scale factor $K(t)$ and the
potential $V(\phi (t))$ from it. This approach was applied in works
\cite{barpar95},\cite{parbar95} as well.

%Nevertheless approximations and numerical investigations are used in a
%large number of inflationary scenarios because of the difficulties in
%obtaining the exact solutions for inflationary models.
%The slow-roll approximation is the most common in use \cite{olive90}.

In  \cite{chezhu95},\cite{chezhu96iv} the exact solutions have been obtained
by taking, first, the scalar factor as the function of time $K=K(t)$
and then determining the evolution of the potential $V=V(t)$ and the
evolution of a scalar field $\phi = \phi (t)$, the dependence between
$V$ and $\phi$ being, in general, parametric one. This approach
called as the method of `fine turning of the potential' \cite{chezhu96iv}.

Another point have been presented in \cite{salbon90}, where exact general
solutions were found for a single scalar field interacting through an
exponential potential in the framework of background field equations
for the Arnowitt-Deser-Misner (ADM) formalism. Approximate analytic
solutions for slowly evolving multiple scalar fields are obtained also.
The Hamilton-Jacobi theory for long-wavelength inhomogeneous universes
are investigated in \cite{salopek91} in the framework of ADM formalism.
Exact inhomogeneous solutions for Yang-Mills field minimally coupled to
gravity have been obtained recently in \cite{szc96}.

In this letter we use the method of `fine turning of the potential'
\cite{chezhu96iv} for obtaining new exact solutions in the case of the
standard inflationary model.
New set of exact solutions for exponential and power law expansion
(inflation) are obtained. The generalised Barrow's solutions \cite{barrow94}
are noted.

It should be mentioned here that the self-interacting scalar field theory
is just the simplest one among the basic models of quantum field theory
\cite{bos92}.
More general approach, as was mentioned above,
is based on the GUT coupled to gravity \cite{svv96iv},\cite{szc96} and can
be reduced in special cases to the chiral inflationary model \cite{ch95iv},
\cite{ch95gc}.

\section{Basic equations and the method}

The inflationary scenario \cite{guth81} as well as its first
modifications \cite{linde82},\cite{albste82} are based on the effective
theory of self-interacting scalar field $\phi$, minimally coupled to
gravity, with the action
\beq{2.1}
S=\int \sr{-g}d^4x \left\{ \frac{R+2\L}{2\k}+
\half \phi_{,i} \phi_{,k} g^{ik} - V(\phi)\right\}.
\eeq
The cosmological constant includes in the action (\ref{2.1}) because of its
importance during the inflationary stage.

Let us consider the standard inflationary model (\ref{2.1})
in the framework of the Friedmann-Robertson-Walker's metric
\beq{2.2}
     dS^2=dt^2-K^2(t)\biggl(\frac{dr^2}{1-\e {r^2}}
     +r^2(d\th^2+\sin^2\th d\vp^2)\biggr).
\eeq
Here $\e=-1,0,+1$ corresponds to open, spatially-flat and closed Universe
respectively.

An important role in inflationary scenarios belongs to the effective
potential of self-interaction $V(\phi)$.
The form of the $V(\phi)$ reflects the physical phenomenon at the very early
Universe: cosmological phase transitions and the symmetry restoration at
high temperatures $T$ in GUTs. It is well known that the form of a
potential
is changed while phase transitions occur and a temperature increases
\cite{linde90},\cite{olive90}. The potential depends on the temperature
and this dependence is due to quantum one-loop corrections in finite temperature
field theory.

Thus the form of the effective scalar potential depends on
the type of a field theory, puting into the physical basis,
and tends to change
when physical phenomenon occur in the development of a cosmological time $t$.

One more restriction on the form of the potential $V(\phi)$ came from the
fine turning procedure \cite{shavil84}. As an example one can consider the
situation with the Coleman-Weinberg potential \cite{colwei73} discussed in
\cite{ch96pla}.

In spite of the restrictions above
we can come to the conclusion, that the form of the
effective potential $V(\phi )$ does not fix and is subjected to change
with the evolution of the Universe.
Puting into basis possible variations of the $V(\phi(t))$ we can
ask a question:"What kind of the $V(\phi(t))$ admits
the exact solutions with an exponential or power law expansion of the FRW
Universe?"
The answer to this question
will give us an explicit form of the
potential which leads to the given rate of the expansion of the Universe.

The system of Einstein's and nonlinear scalar field equations, corresponding
to the model (\ref{2.1}) in the FRW spaces (\ref{2.2}), reads
\beqa{3.1}
  &&\frac{K_{44}}{K}+\frac{2K_{4}^2}{K^2}+\frac{2\epsilon}{K^2}
  =-\Lambda+\kappa{V(\phi)},\\ \label{3.2}
  &&-\frac{3K_{44}}{K}=\Lambda+\kappa(\phi^2_4-{V(\phi)}),\\ \label{3.3}
  &&\phi_{44}+3\frac{K_{4}}{K}\phi_4+\frac{dV(\phi)}{d\phi}=0.
\eeqa
References on a limited number of the exact solutions for
the system (\ref{3.1})-(\ref{3.3}) is given above.

Considering the equations (\ref{3.1})-(\ref{3.3}) one can find that
the last equation (\ref{3.3}) is the differential consequence of the
(\ref{3.1}) and (\ref{3.2}).
Therefore the following analysis will use just Einstein's equations
(\ref{3.1}) and (\ref{3.2}), which can be reduced to the form
where the functions
$V(t)\equiv{V(\phi(t))}$ and $\phi(t)$
are expressed through the function
$K(t)$ and their derivatives:

\beqa{3.6}
  &&V(t)=\frac{1}{\kappa}\left(\Lambda+\frac{K_{44}}{K}
  +\frac{2K_{4}^2}{K^2}+\frac{2\epsilon}{K^2}\right),\\
\label{3.7}
 &&\phi(t)=\pm\sqrt{\frac{2}{\kappa}}\int\left(\sqrt{-\frac{d^2\ln{K}}{dt^2}%
 +\frac{\epsilon}{K^2}}\right)dt+\phi_0,
\eeqa
In (\ref{3.7}) $\phi_0$ is a constant of integration.

By giving the rate of the expansion as the function for a scale factor
 on time $K=K(t)$ , we can find the functions
$\phi(t)$ and $V(t)$ which are necessary for chosen type of the Universe's
evolution.
It is obvious, that the pair of the function (\ref{3.6}) and (\ref{3.7})
gives the parametric dependence
$V=V(\phi)$.
In some cases, after calculation of the right hand sides in
 (\ref{3.6}),
(\ref{3.7}), it is possible to find the explicit dependence
$V=V(\phi)$ by eliminating $t$.

\section{Exponential type solutions}

Let us consider the case when the scale factor $K(t)$ of the Universe
grows up very fast by the exponential type law.

We have to mention about
the simplest solutions of the exponential type in the form
$K(t)=K_0e^{\lambda{t}}$ , where $K_0$ and $\l$ are constants.

\begin{itemize}
\item For the spatially-flat Universe ($\e=0$) one can find \\
$\phi=\phi_0=const;~V(\phi)=V_0=const;~\L=0 $ or $ \L \ne 0 $.\\
This solution corresponds to slow-roll regime \cite{linde90}.
\item Closed Universe $(\e=1)$ also admits an exponential expansion
if the potential takes the form \\
$V(\phi)=\frac{1}{\kappa}(\Lambda+3\lambda^2+\frac{2\e\lambda^2}{K_0^2q^2}
(\phi-\phi_0)^2)$\\
\end{itemize}

In the chiral inflationary model \cite{ch95iv}, \cite{ch95gc} it was found
the solutions where $K(t)\propto \cosh^\al t ~(\sinh^\al t)$ and
 $K(t)\propto \cos^\al t ~(\sin^\al t)$, with $\al=\frac{1}{3}$ or $\al=1$.
We will investigate the possibilities to obtain the same rate of expansion
in standard inflationary model (\ref{2.1}).
Therefore we choose the scalar factor of the Universe in the form
\beq{3.8}
    K(t)=K_0\cdot{\rm ch^\al\{\lambda{t}\}}
\eeq
or
\beq{3.9}
    K(t)=K_0\cdot{\rm sh^\al\{\lambda{t}\}}
\eeq
where $\l$--constant.
The scalar factor (\ref{3.9}) have been obtained for the spatially
flat FRW Universe in \cite{barrow94}.
For the sake of simplicity let $\al=1$.

In the case \rf{3.8} the scale factor $K(t)$ is finite in the initial
moment $t=0$.
In the case \rf{3.9} the scalar factor equals to zero
and we have the initial singularity. In the both cases the solutions for
$\phi(t),V(t)$ and $V(\phi)$ can be obtained explicitly.
Starting from the case \rf{3.8} one can find
\beqa{3.10} \nonumber
&&\phi(t)=\mp\frac{q}{\lambda}\arctg\lbrace\frac{1}{{\rm sh}(\lambda{t})}\rbrace,\\
&&V(\phi)=\frac{1}{\kappa}\left[\Lambda+3\lambda^2-q^2\kappa^2\cos^2\frac{2\lambda\phi}{q}\right],
\eeqa
where {\large $q=\frac{1}{\kappa}\sqrt{\frac{2\e}{K_0^2}-2\lambda^2}$}.
In the c se of (\ref{3.10}) the dependence $V$ on the field $\phi$ is
the periodic one (see fig. 1 and 2 ).
The evolution of the Universe begins from finite radios which
then exponentially expands to infinite one.
This situation corresponds to a rolling from unstable equilibrium points
of the maxima
to the stable equilibrium points of the potential's minimum on the
plane $V-\phi$.

In the case \rf{3.9} the solution can be presented as
\beqa{3.11} \nonumber
&&\phi(t)=\pm\frac{q}{2\lambda}{\rm ln}\frac{{\rm áh}(\lambda{t})-1}{{\rm ch}(\lambda{t})+1},\\
&&V(\phi)=\frac{1}{\kappa}\left[\Lambda+3\lambda^2+q^2\kappa^2{\rm sh}^2\frac{2\lambda\phi}{q}\right],
\eeqa
where {\large $q=\frac{1}{\kappa}\sqrt{\frac{2\e  }{K_0^2}+2\lambda^2}$.}
The Universe evaluates from the singular state, corresponding to infinity
large value of the field $\phi$, to the state with infinitely large radius,
which corresponds to the single local minimum of the potential
$V(\phi)$.
These solutions coinside with Barrow's one \cite{barrow94} when $\e   =0$.

Let us mention also that the potential (\ref{3.11}) leads to one more exact
solution, corresponding to harmonic variation of the scale factor
$$
K(t)=a_1\sin\{\lambda_1t\}
$$
where $a_1$and $\l_1$ are special values of parameters.
The solution is described by
\beqa{3.12} \nonumber
&&\phi(t)=\pm\frac{q}{2\lambda_1}{\rm ln}\frac{1-{\rm cos}(\lambda_1{t})}{1+{\rm cos}(\lambda_1{t})},\\
&&V(\phi)=\frac{1}{\kappa}\left[\Lambda-3\lambda_1^2+q_1^2\kappa^2{\rm ch}^2\frac{2\lambda_1\phi}{q_1}\right],
\eeqa
where {\large $q_1=\frac{1}{\kappa}\sqrt{\frac{2\e  }{a_1^2}+2\lambda_1^2}$.}
This type of solutions called in \cite{barrow94} as a trigonometric
counterpart to the solution \rf{3.9}.

The potentials \rf{3.11} and \rf{3.12} coincide with the additive constant
accuracy if
$
     \lambda_1=\lambda, \quad q_1=q.
$
The last regime  \rf{3.12} corresponds to periodical crossing of the singular
state.

Let us mention now that all solutions presented in this sections
exist independent off the type
of the Universe: open, spatially-flat or closed.
The type of the Universe depends on the sign of subradical's expression
$\frac{2\e}{a_1^2}\pm{2}\lambda_1^2 $.
If $\frac{2\e}{a^2}\pm{2}\lambda^2 > 0$, then the above cases are realized.
If  $\frac{2\e}{a^2}\pm{2}\lambda^2 < 0$, then the cases
\rf{3.10} and \rf{3.11} are changed by places.
The exceptional case $q=0$ corresponds to zero or any constant value of the
field.

Let us note again that the solutions \rf{3.11} and \rf{3.12} can
be considered as the generalisation of Barrow's solutions \cite{barrow94}
for the case of open and closed universes.

\section{Power law inflation}
For the power law inflation the scale factor can be
presented in the form
$$
K(t)=K_0t^m.
$$
The integral in the right hand side of
(\ref{3.7}) can be calculated explicitly.
Therefore we can find  $(m\not=1)$
\begin{eqnarray} \label{EqVt}
  &&V(\phi(t))=\frac{m}{\kappa}\left(\frac{\Lambda}{m} - t^{-2}(3m-1-2\alpha{t}^{-2m+2})\right)\\
\label{EqPhit}
 &&\phi(t)=\pm\sqrt{\frac{1}{2\kappa}}\frac{m}{1-m}\left\{2\sqrt{1+\alpha{t}^{-2m+2}}%
 +{\rm ln}\left(\frac{\sqrt{1+\alpha{t}^{-2m+2}}+1}{\sqrt{1+\alpha{t}^{-2m+2}}-1}\right)\right\}+\phi_0,
\end{eqnarray}
where $\alpha=\e {K_0^{-2}}/m$.
%\vfill\eject
\begin{itemize}
\item As was mentioned in \cite{ch96pla} for the spatially-flat
Universe ($\e=0$) the solution for arbitrary $m$ has the form
\begin{equation}
\phi= \pm \sr{2m/\k} \ln t +\phi_0,~ V(t)=\k^{-1}(\L+(m+3m^2)t^{-2}).
\end{equation}
Eliminating $t$ we find an exponential dependence $V$ on $\phi$
\begin{equation}
V(\phi)=\k^{-1}\{\L+(m+3m^2) \exp (-\sr{2\k m^{-1}}\{\phi-\phi_0\})\},
\end{equation}
what is usually the definition of the power law inflation.
\item
In the case of open and closed Universe $(\e \ne 0)$ it is possible to
find an explicit dependence $V$ on $\phi$ just for some values of $m$.
For example, if $m=1$ (in this case the formulas
\rf{EqVt} and \rf{EqPhit} do not work)
\begin{equation}
V(\phi)=\L/\k +\k^{-1} \exp \{- {2(\phi -\phi_0) \over \pm \sr{2/\k}
\sr{1+\ve K_0^{-2}}}\},
\end{equation}

\item General behaviour all functions $K(t),~V(\phi),~ \phi (t)$
are common for any
$m > 1$. The figure 3 illustrates
the behaviour of the functions in the case $m=5$.
\end{itemize}
\section{Conclusions}
The idea that the shape of the potential $V(\phi)$ in cosmological
inflationary models does not fixed allows us to find new exact solutions
for exponential type inflation. In the case of power law inflation it was
found the general solution. Both cases are considered for FRW Universes.
The following task will be to find an appropriate theory for obtaining the
relevant shape of the effective potential $V(\phi)$.

\section*{Acknowledgements}

We thankful to anonymous refree for informing us about the articles which
are close to our work. We also grateful to J.D.Barrow and D.S.Salopek
for sending preprints on relevant topic.

This work was partly supported by the CCPP `Cosmion' in the project on
CosmoParticle Physics.

\vfill\eject

%\vfill\eject
\section*{Figure captions:}
We use a relative units for time $t$,the scalar field $\phi$, the potential
$V(\phi)$ and scalar factor $K$ for the sake to illustrate the general
behaviour of all functions.
\newpage

\begin{figure} \label{Pic1}
\begin{picture}(145.0,90.0) 
\put(0,0){\line(1,0){145.0}}
\put(0,90.0){\line(1,0){145.0}}
\put(0,0){\line(0,1){90.0}}
\put(145.0,0){\line(0,1){90.0}}
\put(10.0,10.0){\vector(0,1){75.0}}
\put(10.0,10.0){\vector(1,0){120.0}}
\put(13.0,74.0){{\Large $V(\phi)$}}
\put(130.0, 5.0){{\Large $\phi$}}
\put(25.0,80.0){{\large $V(\phi)=\frac{1}{\kappa}(\Lambda+3\lambda^2-\frac{2\ve\lambda^2}{a^2q^2}cos^2\{\frac{2\lambda}{q}(\phi-\phi_0)\})$}}
\put(10.0,10.0){
\put(5.0,28.0){$t=0$}
\put(0.0,25.0){\circle{3.0}}     %61
\put(25.0,8.0){$t=\infty$}
\put(5.0,18.0){\vector(1,-1){5}} %45
\put(30.0,5.0){\circle{3.0}}
\put( 0.0,-0.7){\line(0,1){ 1.4}}
\put(-2.0,-7.0){$0$}
\put(20.0,-0.7){\line(0,1){ 1.4}}
\put(18.0,-7.0){$1$}
\put(40.0,-0.7){\line(0,1){ 1.4}}
\put(38.0,-7.0){$2$}
\put(60.0,-0.7){\line(0,1){ 1.4}}
\put(58.0,-7.0){$3$}
\put(80.0,-0.7){\line(0,1){ 1.4}}
\put(78.0,-7.0){$4$}
\put(100.0,-0.7){\line(0,1){ 1.4}}
\put(98.0,-7.0){$5$}
\put(-0.7, 0.0){\line(1,0){ 1.4}}
\put(-7.0, 0.0){$0$}
\put(-0.7,10.0){\line(1,0){ 1.4}}
\put(-7.0,10.0){$1$}
\put(-0.7,20.0){\line(1,0){ 1.4}}
\put(-7.0,20.0){$2$}
\put(-0.7,30.0){\line(1,0){ 1.4}}
\put(-7.0,30.0){$3$}
\put(-0.7,40.0){\line(1,0){ 1.4}}
\put(-7.0,40.0){$4$}
\put(-0.7,50.0){\line(1,0){ 1.4}}
\put(-7.0,50.0){$5$}
\put(-0.7,60.0){\line(1,0){ 1.4}}
\put(-7.0,60.0){$6$}
\put( 0.0,24.9){\circle*{0.5}}
\put( 0.3,24.9){\circle*{0.5}}
\put( 0.5,24.8){\circle*{0.5}}
\put( 0.8,24.8){\circle*{0.5}}
\put( 1.0,24.7){\circle*{0.5}}
\put( 1.2,24.6){\circle*{0.5}}
\put( 1.5,24.6){\circle*{0.5}}
\put( 1.7,24.5){\circle*{0.5}}
\put( 2.0,24.4){\circle*{0.5}}
\put( 2.3,24.3){\circle*{0.5}}
\put( 2.5,24.2){\circle*{0.5}}
\put( 2.8,24.1){\circle*{0.5}}
\put( 3.0,24.0){\circle*{0.5}}
\put( 3.3,23.9){\circle*{0.5}}
\put( 3.5,23.8){\circle*{0.5}}
\put( 3.8,23.7){\circle*{0.5}}
\put( 4.0,23.6){\circle*{0.5}}
\put( 4.3,23.4){\circle*{0.5}}
\put( 4.5,23.3){\circle*{0.5}}
\put( 4.7,23.2){\circle*{0.5}}
\put( 5.0,23.0){\circle*{0.5}}
\put( 5.3,22.9){\circle*{0.5}}
\put( 5.5,22.7){\circle*{0.5}}
\put( 5.7,22.5){\circle*{0.5}}
\put( 6.0,22.4){\circle*{0.5}}
\put( 6.2,22.2){\circle*{0.5}}
\put( 6.5,22.0){\circle*{0.5}}
\put( 6.8,21.8){\circle*{0.5}}
\put( 7.0,21.7){\circle*{0.5}}
\put( 7.3,21.5){\circle*{0.5}}
\put( 7.5,21.3){\circle*{0.5}}
\put( 7.8,21.1){\circle*{0.5}}
\put( 8.0,20.9){\circle*{0.5}}
\put( 8.2,20.7){\circle*{0.5}}
\put( 8.5,20.5){\circle*{0.5}}
\put( 8.8,20.3){\circle*{0.5}}
\put( 9.0,20.0){\circle*{0.5}}
\put( 9.3,19.8){\circle*{0.5}}
\put( 9.5,19.6){\circle*{0.5}}
\put( 9.8,19.4){\circle*{0.5}}
\put(10.0,19.2){\circle*{0.5}}
\put(10.2,18.9){\circle*{0.5}}
\put(10.5,18.7){\circle*{0.5}}
\put(10.8,18.5){\circle*{0.5}}
\put(11.0,18.2){\circle*{0.5}}
\put(11.2,18.0){\circle*{0.5}}
\put(11.5,17.8){\circle*{0.5}}
\put(11.8,17.5){\circle*{0.5}}
\put(12.0,17.3){\circle*{0.5}}
\put(12.3,17.0){\circle*{0.5}}
\put(12.5,16.8){\circle*{0.5}}
\put(12.7,16.5){\circle*{0.5}}
\put(13.0,16.3){\circle*{0.5}}
\put(13.3,16.0){\circle*{0.5}}
\put(13.5,15.8){\circle*{0.5}}
\put(13.8,15.5){\circle*{0.5}}
\put(14.0,15.3){\circle*{0.5}}
\put(14.3,15.0){\circle*{0.5}}
\put(14.5,14.8){\circle*{0.5}}
\put(14.8,14.5){\circle*{0.5}}
\put(15.0,14.3){\circle*{0.5}}
\put(15.2,14.0){\circle*{0.5}}
\put(15.5,13.8){\circle*{0.5}}
\put(15.8,13.5){\circle*{0.5}}
\put(16.0,13.3){\circle*{0.5}}
\put(16.2,13.1){\circle*{0.5}}
\put(16.5,12.8){\circle*{0.5}}
\put(16.8,12.6){\circle*{0.5}}
\put(17.0,12.3){\circle*{0.5}}
\put(17.3,12.1){\circle*{0.5}}
\put(17.5,11.8){\circle*{0.5}}
\put(17.7,11.6){\circle*{0.5}}
\put(18.0,11.4){\circle*{0.5}}
\put(18.3,11.1){\circle*{0.5}}
\put(18.5,10.9){\circle*{0.5}}
\put(18.8,10.7){\circle*{0.5}}
\put(19.0,10.5){\circle*{0.5}}
\put(19.3,10.2){\circle*{0.5}}
\put(19.5,10.0){\circle*{0.5}}
\put(19.8, 9.8){\circle*{0.5}}
\put(20.0, 9.6){\circle*{0.5}}
\put(20.3, 9.4){\circle*{0.5}}
\put(20.5, 9.2){\circle*{0.5}}
\put(20.8, 9.0){\circle*{0.5}}
\put(21.0, 8.8){\circle*{0.5}}
\put(21.2, 8.6){\circle*{0.5}}
\put(21.5, 8.4){\circle*{0.5}}
\put(21.7, 8.2){\circle*{0.5}}
\put(22.0, 8.0){\circle*{0.5}}
\put(22.3, 7.9){\circle*{0.5}}
\put(22.5, 7.7){\circle*{0.5}}
\put(22.8, 7.5){\circle*{0.5}}
\put(23.0, 7.4){\circle*{0.5}}
\put(23.3, 7.2){\circle*{0.5}}
\put(23.5, 7.0){\circle*{0.5}}
\put(23.8, 6.9){\circle*{0.5}}
\put(24.0, 6.7){\circle*{0.5}}
\put(24.2, 6.6){\circle*{0.5}}
\put(24.5, 6.5){\circle*{0.5}}
\put(24.8, 6.3){\circle*{0.5}}
\put(25.0, 6.2){\circle*{0.5}}
\put(25.3, 6.1){\circle*{0.5}}
\put(25.5, 6.0){\circle*{0.5}}
\put(25.8, 5.9){\circle*{0.5}}
\put(26.0, 5.8){\circle*{0.5}}
\put(26.2, 5.7){\circle*{0.5}}
\put(26.5, 5.6){\circle*{0.5}}
\put(26.7, 5.5){\circle*{0.5}}
\put(27.0, 5.4){\circle*{0.5}}
\put(27.3, 5.4){\circle*{0.5}}
\put(27.5, 5.3){\circle*{0.5}}
\put(27.8, 5.3){\circle*{0.5}}
\put(28.0, 5.2){\circle*{0.5}}
\put(28.3, 5.2){\circle*{0.5}}
\put(28.5, 5.1){\circle*{0.5}}
\put(28.8, 5.1){\circle*{0.5}}
\put(29.0, 5.0){\circle*{0.5}}
\put(29.2, 5.0){\circle*{0.5}}
\put(29.5, 5.0){\circle*{0.5}}
\put(29.8, 5.0){\circle*{0.5}}
\put(30.0, 5.0){\circle*{0.5}}
\put(30.0,0.0){\line(0,1){ 5.0}}
\put(30.0,-6.0){{\large $\phi_0$}}\put(30.3, 5.0){\circle*{0.5}}
\put(30.5, 5.0){\circle*{0.5}}
\put(30.8, 5.0){\circle*{0.5}}
\put(31.0, 5.0){\circle*{0.5}}
\put(31.2, 5.1){\circle*{0.5}}
\put(31.5, 5.1){\circle*{0.5}}
\put(31.7, 5.2){\circle*{0.5}}
\put(32.0, 5.2){\circle*{0.5}}
\put(32.3, 5.3){\circle*{0.5}}
\put(32.5, 5.3){\circle*{0.5}}
\put(32.8, 5.4){\circle*{0.5}}
\put(33.0, 5.4){\circle*{0.5}}
\put(33.3, 5.5){\circle*{0.5}}
\put(33.5, 5.6){\circle*{0.5}}
\put(33.8, 5.7){\circle*{0.5}}
\put(34.0, 5.8){\circle*{0.5}}
\put(34.2, 5.9){\circle*{0.5}}
\put(34.5, 6.0){\circle*{0.5}}
\put(34.8, 6.1){\circle*{0.5}}
\put(35.0, 6.2){\circle*{0.5}}
\put(35.3, 6.3){\circle*{0.5}}
\put(35.5, 6.5){\circle*{0.5}}
\put(35.8, 6.6){\circle*{0.5}}
\put(36.0, 6.7){\circle*{0.5}}
\put(36.2, 6.9){\circle*{0.5}}
\put(36.5, 7.0){\circle*{0.5}}
\put(36.7, 7.2){\circle*{0.5}}
\put(37.0, 7.4){\circle*{0.5}}
\put(37.3, 7.5){\circle*{0.5}}
\put(37.5, 7.7){\circle*{0.5}}
\put(37.8, 7.9){\circle*{0.5}}
\put(38.0, 8.0){\circle*{0.5}}
\put(38.3, 8.2){\circle*{0.5}}
\put(38.5, 8.4){\circle*{0.5}}
\put(38.8, 8.6){\circle*{0.5}}
\put(39.0, 8.8){\circle*{0.5}}
\put(39.2, 9.0){\circle*{0.5}}
\put(39.5, 9.2){\circle*{0.5}}
\put(39.8, 9.4){\circle*{0.5}}
\put(40.0, 9.6){\circle*{0.5}}
\put(40.3, 9.8){\circle*{0.5}}
\put(40.5,10.0){\circle*{0.5}}
\put(40.8,10.2){\circle*{0.5}}
\put(41.0,10.5){\circle*{0.5}}
\put(41.2,10.7){\circle*{0.5}}
\put(41.5,10.9){\circle*{0.5}}
\put(41.8,11.1){\circle*{0.5}}
\put(42.0,11.4){\circle*{0.5}}
\put(42.2,11.6){\circle*{0.5}}
\put(42.5,11.8){\circle*{0.5}}
\put(42.8,12.1){\circle*{0.5}}
\put(43.0,12.3){\circle*{0.5}}
\put(43.3,12.6){\circle*{0.5}}
\put(43.5,12.8){\circle*{0.5}}
\put(43.8,13.1){\circle*{0.5}}
\put(44.0,13.3){\circle*{0.5}}
\put(44.3,13.5){\circle*{0.5}}
\put(44.5,13.8){\circle*{0.5}}
\put(44.7,14.0){\circle*{0.5}}
\put(45.0,14.3){\circle*{0.5}}
\put(45.3,14.5){\circle*{0.5}}
\put(45.5,14.8){\circle*{0.5}}
\put(45.8,15.0){\circle*{0.5}}
\put(46.0,15.3){\circle*{0.5}}
\put(46.2,15.5){\circle*{0.5}}
\put(46.5,15.8){\circle*{0.5}}
\put(46.8,16.0){\circle*{0.5}}
\put(47.0,16.3){\circle*{0.5}}
\put(47.2,16.5){\circle*{0.5}}
\put(47.5,16.8){\circle*{0.5}}
\put(47.8,17.0){\circle*{0.5}}
\put(48.0,17.3){\circle*{0.5}}
\put(48.3,17.5){\circle*{0.5}}
\put(48.5,17.8){\circle*{0.5}}
\put(48.8,18.0){\circle*{0.5}}
\put(49.0,18.2){\circle*{0.5}}
\put(49.3,18.5){\circle*{0.5}}
\put(49.5,18.7){\circle*{0.5}}
\put(49.7,18.9){\circle*{0.5}}
\put(50.0,19.2){\circle*{0.5}}
\put(50.3,19.4){\circle*{0.5}}
\put(50.5,19.6){\circle*{0.5}}
\put(50.8,19.8){\circle*{0.5}}
\put(51.0,20.0){\circle*{0.5}}
\put(51.2,20.3){\circle*{0.5}}
\put(51.5,20.5){\circle*{0.5}}
\put(51.8,20.7){\circle*{0.5}}
\put(52.0,20.9){\circle*{0.5}}
\put(52.2,21.1){\circle*{0.5}}
\put(52.5,21.3){\circle*{0.5}}
\put(52.8,21.5){\circle*{0.5}}
\put(53.0,21.7){\circle*{0.5}}
\put(53.3,21.8){\circle*{0.5}}
\put(53.5,22.0){\circle*{0.5}}
\put(53.8,22.2){\circle*{0.5}}
\put(54.0,22.4){\circle*{0.5}}
\put(54.3,22.5){\circle*{0.5}}
\put(54.5,22.7){\circle*{0.5}}
\put(54.7,22.9){\circle*{0.5}}
\put(55.0,23.0){\circle*{0.5}}
\put(55.3,23.2){\circle*{0.5}}
\put(55.5,23.3){\circle*{0.5}}
\put(55.8,23.4){\circle*{0.5}}
\put(56.0,23.6){\circle*{0.5}}
\put(56.2,23.7){\circle*{0.5}}
\put(56.5,23.8){\circle*{0.5}}
\put(56.8,23.9){\circle*{0.5}}
\put(57.0,24.0){\circle*{0.5}}
\put(57.2,24.1){\circle*{0.5}}
\put(57.5,24.2){\circle*{0.5}}
\put(57.8,24.3){\circle*{0.5}}
\put(58.0,24.4){\circle*{0.5}}
\put(58.3,24.5){\circle*{0.5}}
\put(58.5,24.6){\circle*{0.5}}
\put(58.8,24.6){\circle*{0.5}}
\put(59.0,24.7){\circle*{0.5}}
\put(59.3,24.8){\circle*{0.5}}
\put(59.5,24.8){\circle*{0.5}}
\put(59.7,24.9){\circle*{0.5}}
\put(60.0,24.9){\circle*{0.5}}
\put(60.3,24.9){\circle*{0.5}}
\put(60.5,25.0){\circle*{0.5}}
\put(60.8,25.0){\circle*{0.5}}
\put(61.0,25.0){\circle*{0.5}}
\put(61.2,25.0){\circle*{0.5}}
\put(61.5,25.0){\circle*{0.5}}
\put(61.8,25.0){\circle*{0.5}}
\put(62.0,25.0){\circle*{0.5}}
\put(62.2,25.0){\circle*{0.5}}
\put(62.5,24.9){\circle*{0.5}}
\put(62.8,24.9){\circle*{0.5}}
\put(63.0,24.9){\circle*{0.5}}
\put(63.3,24.8){\circle*{0.5}}
\put(63.5,24.8){\circle*{0.5}}
\put(63.8,24.7){\circle*{0.5}}
\put(64.0,24.7){\circle*{0.5}}
\put(64.3,24.6){\circle*{0.5}}
\put(64.5,24.5){\circle*{0.5}}
\put(64.7,24.4){\circle*{0.5}}
\put(65.0,24.4){\circle*{0.5}}
\put(65.3,24.3){\circle*{0.5}}
\put(65.5,24.2){\circle*{0.5}}
\put(65.8,24.1){\circle*{0.5}}
\put(66.0,24.0){\circle*{0.5}}
\put(66.2,23.9){\circle*{0.5}}
\put(66.5,23.7){\circle*{0.5}}
\put(66.8,23.6){\circle*{0.5}}
\put(67.0,23.5){\circle*{0.5}}
\put(67.2,23.3){\circle*{0.5}}
\put(67.5,23.2){\circle*{0.5}}
\put(67.8,23.1){\circle*{0.5}}
\put(68.0,22.9){\circle*{0.5}}
\put(68.3,22.8){\circle*{0.5}}
\put(68.5,22.6){\circle*{0.5}}
\put(68.8,22.4){\circle*{0.5}}
\put(69.0,22.3){\circle*{0.5}}
\put(69.3,22.1){\circle*{0.5}}
\put(69.5,21.9){\circle*{0.5}}
\put(69.7,21.7){\circle*{0.5}}
\put(70.0,21.5){\circle*{0.5}}
\put(70.3,21.3){\circle*{0.5}}
\put(70.5,21.2){\circle*{0.5}}
\put(70.8,21.0){\circle*{0.5}}
\put(71.0,20.7){\circle*{0.5}}
\put(71.2,20.5){\circle*{0.5}}
\put(71.5,20.3){\circle*{0.5}}
\put(71.8,20.1){\circle*{0.5}}
\put(72.0,19.9){\circle*{0.5}}
\put(72.2,19.7){\circle*{0.5}}
\put(72.5,19.5){\circle*{0.5}}
\put(72.8,19.2){\circle*{0.5}}
\put(73.0,19.0){\circle*{0.5}}
\put(73.3,18.8){\circle*{0.5}}
\put(73.5,18.5){\circle*{0.5}}
\put(73.8,18.3){\circle*{0.5}}
\put(74.0,18.1){\circle*{0.5}}
\put(74.3,17.8){\circle*{0.5}}
\put(74.5,17.6){\circle*{0.5}}
\put(74.7,17.4){\circle*{0.5}}
\put(75.0,17.1){\circle*{0.5}}
\put(75.3,16.9){\circle*{0.5}}
\put(75.5,16.6){\circle*{0.5}}
\put(75.8,16.4){\circle*{0.5}}
\put(76.0,16.1){\circle*{0.5}}
\put(76.2,15.9){\circle*{0.5}}
\put(76.5,15.6){\circle*{0.5}}
\put(76.8,15.4){\circle*{0.5}}
\put(77.0,15.1){\circle*{0.5}}
\put(77.2,14.9){\circle*{0.5}}
\put(77.5,14.6){\circle*{0.5}}
\put(77.8,14.4){\circle*{0.5}}
\put(78.0,14.1){\circle*{0.5}}
\put(78.3,13.9){\circle*{0.5}}
\put(78.5,13.6){\circle*{0.5}}
\put(78.8,13.4){\circle*{0.5}}
\put(79.0,13.1){\circle*{0.5}}
\put(79.3,12.9){\circle*{0.5}}
\put(79.5,12.6){\circle*{0.5}}
\put(79.7,12.4){\circle*{0.5}}
\put(80.0,12.2){\circle*{0.5}}
\put(80.3,11.9){\circle*{0.5}}
\put(80.5,11.7){\circle*{0.5}}
\put(80.7,11.5){\circle*{0.5}}
\put(81.0,11.2){\circle*{0.5}}
\put(81.2,11.0){\circle*{0.5}}
\put(81.5,10.8){\circle*{0.5}}
\put(81.8,10.5){\circle*{0.5}}
\put(82.0,10.3){\circle*{0.5}}
\put(82.3,10.1){\circle*{0.5}}
\put(82.5, 9.9){\circle*{0.5}}
\put(82.8, 9.7){\circle*{0.5}}
\put(83.0, 9.5){\circle*{0.5}}
\put(83.2, 9.2){\circle*{0.5}}
\put(83.5, 9.0){\circle*{0.5}}
\put(83.8, 8.8){\circle*{0.5}}
\put(84.0, 8.7){\circle*{0.5}}
\put(84.3, 8.5){\circle*{0.5}}
\put(84.5, 8.3){\circle*{0.5}}
\put(84.8, 8.1){\circle*{0.5}}
\put(85.0, 7.9){\circle*{0.5}}
\put(85.3, 7.7){\circle*{0.5}}
\put(85.5, 7.6){\circle*{0.5}}
\put(85.7, 7.4){\circle*{0.5}}
\put(86.0, 7.2){\circle*{0.5}}
\put(86.2, 7.1){\circle*{0.5}}
\put(86.5, 6.9){\circle*{0.5}}
\put(86.8, 6.8){\circle*{0.5}}
\put(87.0, 6.7){\circle*{0.5}}
\put(87.3, 6.5){\circle*{0.5}}
\put(87.5, 6.4){\circle*{0.5}}
\put(87.8, 6.3){\circle*{0.5}}
\put(88.0, 6.1){\circle*{0.5}}
\put(88.2, 6.0){\circle*{0.5}}
\put(88.5, 5.9){\circle*{0.5}}
\put(88.8, 5.8){\circle*{0.5}}
\put(89.0, 5.7){\circle*{0.5}}
\put(89.3, 5.6){\circle*{0.5}}
\put(89.5, 5.5){\circle*{0.5}}
\put(89.8, 5.5){\circle*{0.5}}
\put(90.0, 5.4){\circle*{0.5}}
\put(90.3, 5.3){\circle*{0.5}}
\put(90.5, 5.3){\circle*{0.5}}
\put(90.7, 5.2){\circle*{0.5}}
\put(91.0, 5.2){\circle*{0.5}}
\put(91.2, 5.1){\circle*{0.5}}
\put(91.5, 5.1){\circle*{0.5}}
\put(91.8, 5.1){\circle*{0.5}}
\put(92.0, 5.0){\circle*{0.5}}
\put(92.3, 5.0){\circle*{0.5}}
\put(92.5, 5.0){\circle*{0.5}}
\put(92.8, 5.0){\circle*{0.5}}
\put(93.0, 5.0){\circle*{0.5}}
\put(93.2, 5.0){\circle*{0.5}}
\put(93.5, 5.0){\circle*{0.5}}
\put(93.8, 5.0){\circle*{0.5}}
\put(94.0, 5.1){\circle*{0.5}}
\put(94.3, 5.1){\circle*{0.5}}
\put(94.5, 5.1){\circle*{0.5}}
\put(94.8, 5.2){\circle*{0.5}}
\put(95.0, 5.2){\circle*{0.5}}
\put(95.3, 5.3){\circle*{0.5}}
\put(95.5, 5.4){\circle*{0.5}}
\put(95.7, 5.4){\circle*{0.5}}
\put(96.0, 5.5){\circle*{0.5}}
\put(96.2, 5.6){\circle*{0.5}}
\put(96.5, 5.7){\circle*{0.5}}
\put(96.8, 5.8){\circle*{0.5}}
\put(97.0, 5.9){\circle*{0.5}}
\put(97.3, 6.0){\circle*{0.5}}
\put(97.5, 6.1){\circle*{0.5}}
\put(97.8, 6.2){\circle*{0.5}}
\put(98.0, 6.3){\circle*{0.5}}
\put(98.2, 6.4){\circle*{0.5}}
\put(98.5, 6.6){\circle*{0.5}}
\put(98.8, 6.7){\circle*{0.5}}
\put(99.0, 6.8){\circle*{0.5}}
\put(99.3, 7.0){\circle*{0.5}}
\put(99.5, 7.1){\circle*{0.5}}
\put(99.8, 7.3){\circle*{0.5}}
\put(100.0, 7.5){\circle*{0.5}}
}
\end{picture}
\caption{The shape of the periodic potential.}
\end{figure}
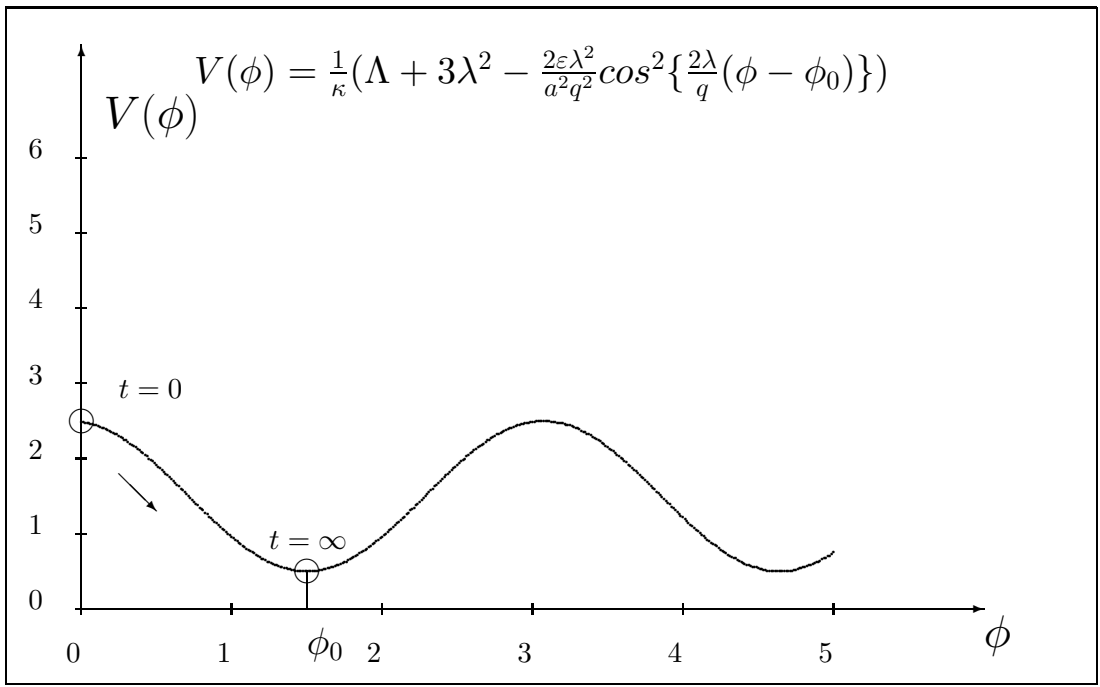

%\vspace{0.5cm}

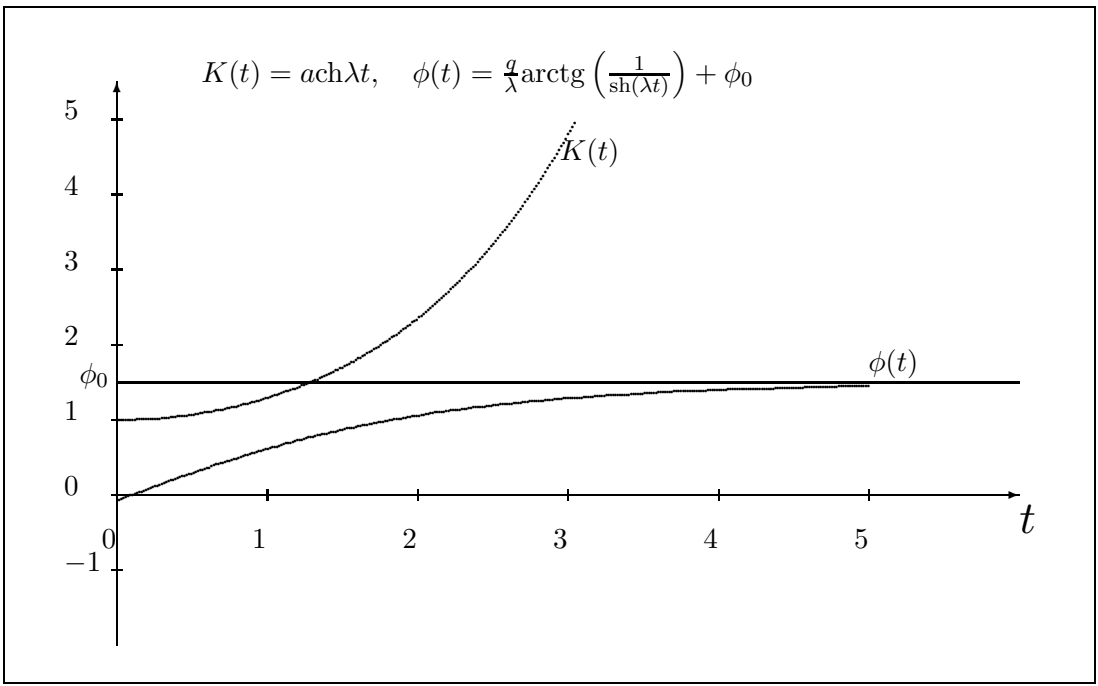
\begin{figure}  \label{Pic2}
\begin{picture}(145.0,90.0)
\put(0,0){\line(1,0){145.0}}
\put(0,90.0){\line(1,0){145.0}}
\put(0,0){\line(0,1){90.0}}
\put(145.0,0){\line(0,1){90.0}}
\put(15.0, 5.0){\vector(0,1){75.0}}
\put(15.0,25.0){\vector(1,0){120.0}}
\put(18.0,74.0){{\Large $\ $}}
\put(135.0,20.0){{\Large $t$}}
\put(25.0,80.0){{ $K(t)=a{\rm ch}{\lambda{t}}, \quad \phi(t)=\frac{q}{\lambda}{\rm arctg}\left(\frac{1}{{\rm sh}(\lambda{t})}\right)+\phi_0$}}
\put(15.0,25.0){
\put( 0.0,-0.7){\line(0,1){ 1.4}}
\put(-2.0,-7.0){$0$}
\put(20.0,-0.7){\line(0,1){ 1.4}}
\put(18.0,-7.0){$1$}
\put(40.0,-0.7){\line(0,1){ 1.4}}
\put(38.0,-7.0){$2$}
\put(60.0,-0.7){\line(0,1){ 1.4}}
\put(58.0,-7.0){$3$}
\put(80.0,-0.7){\line(0,1){ 1.4}}
\put(78.0,-7.0){$4$}
\put(100.0,-0.7){\line(0,1){ 1.4}}
\put(98.0,-7.0){$5$}
\put(-0.7,-10.0){\line(1,0){ 1.4}}
\put(-7.0,-10.0){$-1$}
\put(-0.7, 0.0){\line(1,0){ 1.4}}
\put(-7.0, 0.0){$0$}
\put(-0.7,10.0){\line(1,0){ 1.4}}
\put(-7.0,10.0){$1$}
\put(-0.7,20.0){\line(1,0){ 1.4}}
\put(-7.0,20.0){$2$}
\put(-0.7,30.0){\line(1,0){ 1.4}}
\put(-7.0,30.0){$3$}
\put(-0.7,40.0){\line(1,0){ 1.4}}
\put(-7.0,40.0){$4$}
\put(-0.7,50.0){\line(1,0){ 1.4}}
\put(-7.0,50.0){$5$}
\put( 0.0,15.0){\line(1,0){120.0}}\put(-5.0,15.0){$\phi_0$}\put( 0.0,-0.0){\circle*{0.5}}
\put( 0.0, 0.0){\circle*{0.5}}
\put( 0.3,-0.6){\circle*{0.5}}
\put( 0.3,10.0){\circle*{0.5}}
\put( 0.6,-0.5){\circle*{0.5}}
\put( 0.6,10.0){\circle*{0.5}}
\put( 0.9,-0.4){\circle*{0.5}}
\put( 0.9,10.0){\circle*{0.5}}
\put( 1.2,-0.3){\circle*{0.5}}
\put( 1.2,10.0){\circle*{0.5}}
\put( 1.5,-0.1){\circle*{0.5}}
\put( 1.5,10.0){\circle*{0.5}}
\put( 1.8,-0.0){\circle*{0.5}}
\put( 1.8,10.0){\circle*{0.5}}
\put( 2.1, 0.1){\circle*{0.5}}
\put( 2.1,10.0){\circle*{0.5}}
\put( 2.4, 0.2){\circle*{0.5}}
\put( 2.4,10.0){\circle*{0.5}}
\put( 2.7, 0.3){\circle*{0.5}}
\put( 2.7,10.1){\circle*{0.5}}
\put( 3.0, 0.4){\circle*{0.5}}
\put( 3.0,10.1){\circle*{0.5}}
\put( 3.3, 0.5){\circle*{0.5}}
\put( 3.3,10.1){\circle*{0.5}}
\put( 3.6, 0.6){\circle*{0.5}}
\put( 3.6,10.1){\circle*{0.5}}
\put( 3.9, 0.7){\circle*{0.5}}
\put( 3.9,10.1){\circle*{0.5}}
\put( 4.2, 0.9){\circle*{0.5}}
\put( 4.2,10.1){\circle*{0.5}}
\put( 4.5, 1.0){\circle*{0.5}}
\put( 4.5,10.1){\circle*{0.5}}
\put( 4.8, 1.1){\circle*{0.5}}
\put( 4.8,10.2){\circle*{0.5}}
\put( 5.1, 1.2){\circle*{0.5}}
\put( 5.1,10.2){\circle*{0.5}}
\put( 5.4, 1.3){\circle*{0.5}}
\put( 5.4,10.2){\circle*{0.5}}
\put( 5.7, 1.4){\circle*{0.5}}
\put( 5.7,10.2){\circle*{0.5}}
\put( 6.0, 1.5){\circle*{0.5}}
\put( 6.0,10.3){\circle*{0.5}}
\put( 6.3, 1.6){\circle*{0.5}}
\put( 6.3,10.3){\circle*{0.5}}
\put( 6.6, 1.7){\circle*{0.5}}
\put( 6.6,10.3){\circle*{0.5}}
\put( 6.9, 1.9){\circle*{0.5}}
\put( 6.9,10.3){\circle*{0.5}}
\put( 7.2, 2.0){\circle*{0.5}}
\put( 7.2,10.4){\circle*{0.5}}
\put( 7.5, 2.1){\circle*{0.5}}
\put( 7.5,10.4){\circle*{0.5}}
\put( 7.8, 2.2){\circle*{0.5}}
\put( 7.8,10.4){\circle*{0.5}}
\put( 8.1, 2.3){\circle*{0.5}}
\put( 8.1,10.5){\circle*{0.5}}
\put( 8.4, 2.4){\circle*{0.5}}
\put( 8.4,10.5){\circle*{0.5}}
\put( 8.7, 2.5){\circle*{0.5}}
\put( 8.7,10.5){\circle*{0.5}}
\put( 9.0, 2.6){\circle*{0.5}}
\put( 9.0,10.6){\circle*{0.5}}
\put( 9.3, 2.7){\circle*{0.5}}
\put( 9.3,10.6){\circle*{0.5}}
\put( 9.6, 2.8){\circle*{0.5}}
\put( 9.6,10.7){\circle*{0.5}}
\put( 9.9, 2.9){\circle*{0.5}}
\put( 9.9,10.7){\circle*{0.5}}
\put(10.2, 3.0){\circle*{0.5}}
\put(10.2,10.7){\circle*{0.5}}
\put(10.5, 3.1){\circle*{0.5}}
\put(10.5,10.8){\circle*{0.5}}
\put(10.8, 3.2){\circle*{0.5}}
\put(10.8,10.8){\circle*{0.5}}
\put(11.1, 3.3){\circle*{0.5}}
\put(11.1,10.9){\circle*{0.5}}
\put(11.4, 3.4){\circle*{0.5}}
\put(11.4,10.9){\circle*{0.5}}
\put(11.7, 3.5){\circle*{0.5}}
\put(11.7,11.0){\circle*{0.5}}
\put(12.0, 3.6){\circle*{0.5}}
\put(12.0,11.0){\circle*{0.5}}
\put(12.3, 3.7){\circle*{0.5}}
\put(12.3,11.1){\circle*{0.5}}
\put(12.6, 3.9){\circle*{0.5}}
\put(12.6,11.1){\circle*{0.5}}
\put(12.9, 4.0){\circle*{0.5}}
\put(12.9,11.2){\circle*{0.5}}
\put(13.2, 4.1){\circle*{0.5}}
\put(13.2,11.3){\circle*{0.5}}
\put(13.5, 4.2){\circle*{0.5}}
\put(13.5,11.3){\circle*{0.5}}
\put(13.8, 4.3){\circle*{0.5}}
\put(13.8,11.4){\circle*{0.5}}
\put(14.1, 4.3){\circle*{0.5}}
\put(14.1,11.4){\circle*{0.5}}
\put(14.4, 4.4){\circle*{0.5}}
\put(14.4,11.5){\circle*{0.5}}
\put(14.7, 4.5){\circle*{0.5}}
\put(14.7,11.6){\circle*{0.5}}
\put(15.0, 4.6){\circle*{0.5}}
\put(15.0,11.6){\circle*{0.5}}
\put(15.3, 4.7){\circle*{0.5}}
\put(15.3,11.7){\circle*{0.5}}
\put(15.6, 4.8){\circle*{0.5}}
\put(15.6,11.8){\circle*{0.5}}
\put(15.9, 4.9){\circle*{0.5}}
\put(15.9,11.8){\circle*{0.5}}
\put(16.2, 5.0){\circle*{0.5}}
\put(16.2,11.9){\circle*{0.5}}
\put(16.5, 5.1){\circle*{0.5}}
\put(16.5,12.0){\circle*{0.5}}
\put(16.8, 5.2){\circle*{0.5}}
\put(16.8,12.1){\circle*{0.5}}
\put(17.1, 5.3){\circle*{0.5}}
\put(17.1,12.1){\circle*{0.5}}
\put(17.4, 5.4){\circle*{0.5}}
\put(17.4,12.2){\circle*{0.5}}
\put(17.7, 5.5){\circle*{0.5}}
\put(17.7,12.3){\circle*{0.5}}
\put(18.0, 5.6){\circle*{0.5}}
\put(18.0,12.4){\circle*{0.5}}
\put(18.3, 5.7){\circle*{0.5}}
\put(18.3,12.4){\circle*{0.5}}
\put(18.6, 5.8){\circle*{0.5}}
\put(18.6,12.5){\circle*{0.5}}
\put(18.9, 5.9){\circle*{0.5}}
\put(18.9,12.6){\circle*{0.5}}
\put(19.2, 5.9){\circle*{0.5}}
\put(19.2,12.7){\circle*{0.5}}
\put(19.5, 6.0){\circle*{0.5}}
\put(19.5,12.8){\circle*{0.5}}
\put(19.8, 6.1){\circle*{0.5}}
\put(19.8,12.9){\circle*{0.5}}
\put(20.1, 6.2){\circle*{0.5}}
\put(20.1,13.0){\circle*{0.5}}
\put(20.4, 6.3){\circle*{0.5}}
\put(20.4,13.1){\circle*{0.5}}
\put(20.7, 6.4){\circle*{0.5}}
\put(20.7,13.2){\circle*{0.5}}
\put(21.0, 6.5){\circle*{0.5}}
\put(21.0,13.3){\circle*{0.5}}
\put(21.3, 6.5){\circle*{0.5}}
\put(21.3,13.4){\circle*{0.5}}
\put(21.6, 6.6){\circle*{0.5}}
\put(21.6,13.5){\circle*{0.5}}
\put(21.9, 6.7){\circle*{0.5}}
\put(21.9,13.6){\circle*{0.5}}
\put(22.2, 6.8){\circle*{0.5}}
\put(22.2,13.7){\circle*{0.5}}
\put(22.5, 6.9){\circle*{0.5}}
\put(22.5,13.8){\circle*{0.5}}
\put(22.8, 7.0){\circle*{0.5}}
\put(22.8,13.9){\circle*{0.5}}
\put(23.1, 7.0){\circle*{0.5}}
\put(23.1,14.0){\circle*{0.5}}
\put(23.4, 7.1){\circle*{0.5}}
\put(23.4,14.1){\circle*{0.5}}
\put(23.7, 7.2){\circle*{0.5}}
\put(23.7,14.2){\circle*{0.5}}
\put(24.0, 7.3){\circle*{0.5}}
\put(24.0,14.3){\circle*{0.5}}
\put(24.3, 7.4){\circle*{0.5}}
\put(24.3,14.4){\circle*{0.5}}
\put(24.6, 7.4){\circle*{0.5}}
\put(24.6,14.6){\circle*{0.5}}
\put(24.9, 7.5){\circle*{0.5}}
\put(24.9,14.7){\circle*{0.5}}
\put(25.2, 7.6){\circle*{0.5}}
\put(25.2,14.8){\circle*{0.5}}
\put(25.5, 7.7){\circle*{0.5}}
\put(25.5,14.9){\circle*{0.5}}
\put(25.8, 7.7){\circle*{0.5}}
\put(25.8,15.1){\circle*{0.5}}
\put(26.1, 7.8){\circle*{0.5}}
\put(26.1,15.2){\circle*{0.5}}
\put(26.4, 7.9){\circle*{0.5}}
\put(26.4,15.3){\circle*{0.5}}
\put(26.7, 8.0){\circle*{0.5}}
\put(26.7,15.4){\circle*{0.5}}
\put(27.0, 8.0){\circle*{0.5}}
\put(27.0,15.6){\circle*{0.5}}
\put(27.3, 8.1){\circle*{0.5}}
\put(27.3,15.7){\circle*{0.5}}
\put(27.6, 8.2){\circle*{0.5}}
\put(27.6,15.9){\circle*{0.5}}
\put(27.9, 8.2){\circle*{0.5}}
\put(27.9,16.0){\circle*{0.5}}
\put(28.2, 8.3){\circle*{0.5}}
\put(28.2,16.1){\circle*{0.5}}
\put(28.5, 8.4){\circle*{0.5}}
\put(28.5,16.3){\circle*{0.5}}
\put(28.8, 8.5){\circle*{0.5}}
\put(28.8,16.4){\circle*{0.5}}
\put(29.1, 8.5){\circle*{0.5}}
\put(29.1,16.6){\circle*{0.5}}
\put(29.4, 8.6){\circle*{0.5}}
\put(29.4,16.7){\circle*{0.5}}
\put(29.7, 8.7){\circle*{0.5}}
\put(29.7,16.9){\circle*{0.5}}
\put(30.0, 8.7){\circle*{0.5}}
\put(30.0,17.0){\circle*{0.5}}
\put(30.3, 8.8){\circle*{0.5}}
\put(30.3,17.2){\circle*{0.5}}
\put(30.6, 8.9){\circle*{0.5}}
\put(30.6,17.3){\circle*{0.5}}
\put(30.9, 8.9){\circle*{0.5}}
\put(30.9,17.5){\circle*{0.5}}
\put(31.2, 9.0){\circle*{0.5}}
\put(31.2,17.7){\circle*{0.5}}
\put(31.5, 9.0){\circle*{0.5}}
\put(31.5,17.8){\circle*{0.5}}
\put(31.8, 9.1){\circle*{0.5}}
\put(31.8,18.0){\circle*{0.5}}
\put(32.1, 9.2){\circle*{0.5}}
\put(32.1,18.2){\circle*{0.5}}
\put(32.4, 9.2){\circle*{0.5}}
\put(32.4,18.3){\circle*{0.5}}
\put(32.7, 9.3){\circle*{0.5}}
\put(32.7,18.5){\circle*{0.5}}
\put(33.0, 9.4){\circle*{0.5}}
\put(33.0,18.7){\circle*{0.5}}
\put(33.3, 9.4){\circle*{0.5}}
\put(33.3,18.9){\circle*{0.5}}
\put(33.6, 9.5){\circle*{0.5}}
\put(33.6,19.0){\circle*{0.5}}
\put(33.9, 9.5){\circle*{0.5}}
\put(33.9,19.2){\circle*{0.5}}
\put(34.2, 9.6){\circle*{0.5}}
\put(34.2,19.4){\circle*{0.5}}
\put(34.5, 9.6){\circle*{0.5}}
\put(34.5,19.6){\circle*{0.5}}
\put(34.8, 9.7){\circle*{0.5}}
\put(34.8,19.8){\circle*{0.5}}
\put(35.1, 9.8){\circle*{0.5}}
\put(35.1,20.0){\circle*{0.5}}
\put(35.4, 9.8){\circle*{0.5}}
\put(35.4,20.2){\circle*{0.5}}
\put(35.7, 9.9){\circle*{0.5}}
\put(35.7,20.4){\circle*{0.5}}
\put(36.0, 9.9){\circle*{0.5}}
\put(36.0,20.6){\circle*{0.5}}
\put(36.3,10.0){\circle*{0.5}}
\put(36.3,20.8){\circle*{0.5}}
\put(36.6,10.0){\circle*{0.5}}
\put(36.6,21.0){\circle*{0.5}}
\put(36.9,10.1){\circle*{0.5}}
\put(36.9,21.2){\circle*{0.5}}
\put(37.2,10.1){\circle*{0.5}}
\put(37.2,21.4){\circle*{0.5}}
\put(37.5,10.2){\circle*{0.5}}
\put(37.5,21.6){\circle*{0.5}}
\put(37.8,10.2){\circle*{0.5}}
\put(37.8,21.8){\circle*{0.5}}
\put(38.1,10.3){\circle*{0.5}}
\put(38.1,22.1){\circle*{0.5}}
\put(38.4,10.3){\circle*{0.5}}
\put(38.4,22.3){\circle*{0.5}}
\put(38.7,10.4){\circle*{0.5}}
\put(38.7,22.5){\circle*{0.5}}
\put(39.0,10.4){\circle*{0.5}}
\put(39.0,22.7){\circle*{0.5}}
\put(39.3,10.5){\circle*{0.5}}
\put(39.3,23.0){\circle*{0.5}}
\put(39.6,10.5){\circle*{0.5}}
\put(39.6,23.2){\circle*{0.5}}
\put(39.9,10.6){\circle*{0.5}}
\put(39.9,23.4){\circle*{0.5}}
\put(40.2,10.6){\circle*{0.5}}
\put(40.2,23.7){\circle*{0.5}}
\put(40.5,10.7){\circle*{0.5}}
\put(40.5,23.9){\circle*{0.5}}
\put(40.8,10.7){\circle*{0.5}}
\put(40.8,24.2){\circle*{0.5}}
\put(41.1,10.8){\circle*{0.5}}
\put(41.1,24.4){\circle*{0.5}}
\put(41.4,10.8){\circle*{0.5}}
\put(41.4,24.7){\circle*{0.5}}
\put(41.7,10.9){\circle*{0.5}}
\put(41.7,24.9){\circle*{0.5}}
\put(42.0,10.9){\circle*{0.5}}
\put(42.0,25.2){\circle*{0.5}}
\put(42.3,11.0){\circle*{0.5}}
\put(42.3,25.5){\circle*{0.5}}
\put(42.6,11.0){\circle*{0.5}}
\put(42.6,25.7){\circle*{0.5}}
\put(42.9,11.0){\circle*{0.5}}
\put(42.9,26.0){\circle*{0.5}}
\put(43.2,11.1){\circle*{0.5}}
\put(43.2,26.3){\circle*{0.5}}
\put(43.5,11.1){\circle*{0.5}}
\put(43.5,26.5){\circle*{0.5}}
\put(43.8,11.2){\circle*{0.5}}
\put(43.8,26.8){\circle*{0.5}}
\put(44.1,11.2){\circle*{0.5}}
\put(44.1,27.1){\circle*{0.5}}
\put(44.4,11.3){\circle*{0.5}}
\put(44.4,27.4){\circle*{0.5}}
\put(44.7,11.3){\circle*{0.5}}
\put(44.7,27.7){\circle*{0.5}}
\put(45.0,11.3){\circle*{0.5}}
\put(45.0,28.0){\circle*{0.5}}
\put(45.3,11.4){\circle*{0.5}}
\put(45.3,28.3){\circle*{0.5}}
\put(45.6,11.4){\circle*{0.5}}
\put(45.6,28.5){\circle*{0.5}}
\put(45.9,11.5){\circle*{0.5}}
\put(45.9,28.9){\circle*{0.5}}
\put(46.2,11.5){\circle*{0.5}}
\put(46.2,29.2){\circle*{0.5}}
\put(46.5,11.5){\circle*{0.5}}
\put(46.5,29.5){\circle*{0.5}}
\put(46.8,11.6){\circle*{0.5}}
\put(46.8,29.8){\circle*{0.5}}
\put(47.1,11.6){\circle*{0.5}}
\put(47.1,30.1){\circle*{0.5}}
\put(47.4,11.7){\circle*{0.5}}
\put(47.4,30.4){\circle*{0.5}}
\put(47.7,11.7){\circle*{0.5}}
\put(47.7,30.7){\circle*{0.5}}
\put(48.0,11.7){\circle*{0.5}}
\put(48.0,31.1){\circle*{0.5}}
\put(48.3,11.8){\circle*{0.5}}
\put(48.3,31.4){\circle*{0.5}}
\put(48.6,11.8){\circle*{0.5}}
\put(48.6,31.7){\circle*{0.5}}
\put(48.9,11.8){\circle*{0.5}}
\put(48.9,32.1){\circle*{0.5}}
\put(49.2,11.9){\circle*{0.5}}
\put(49.2,32.4){\circle*{0.5}}
\put(49.5,11.9){\circle*{0.5}}
\put(49.5,32.8){\circle*{0.5}}
\put(49.8,11.9){\circle*{0.5}}
\put(49.8,33.1){\circle*{0.5}}
\put(50.1,12.0){\circle*{0.5}}
\put(50.1,33.5){\circle*{0.5}}
\put(50.4,12.0){\circle*{0.5}}
\put(50.4,33.9){\circle*{0.5}}
\put(50.7,12.0){\circle*{0.5}}
\put(50.7,34.2){\circle*{0.5}}
\put(51.0,12.1){\circle*{0.5}}
\put(51.0,34.6){\circle*{0.5}}
\put(51.3,12.1){\circle*{0.5}}
\put(51.3,35.0){\circle*{0.5}}
\put(51.6,12.1){\circle*{0.5}}
\put(51.6,35.3){\circle*{0.5}}
\put(51.9,12.2){\circle*{0.5}}
\put(51.9,35.7){\circle*{0.5}}
\put(52.2,12.2){\circle*{0.5}}
\put(52.2,36.1){\circle*{0.5}}
\put(52.5,12.2){\circle*{0.5}}
\put(52.5,36.5){\circle*{0.5}}
\put(52.8,12.3){\circle*{0.5}}
\put(52.8,36.9){\circle*{0.5}}
\put(53.1,12.3){\circle*{0.5}}
\put(53.1,37.3){\circle*{0.5}}
\put(53.4,12.3){\circle*{0.5}}
\put(53.4,37.7){\circle*{0.5}}
\put(53.7,12.3){\circle*{0.5}}
\put(53.7,38.1){\circle*{0.5}}
\put(54.0,12.4){\circle*{0.5}}
\put(54.0,38.5){\circle*{0.5}}
\put(54.3,12.4){\circle*{0.5}}
\put(54.3,39.0){\circle*{0.5}}
\put(54.6,12.4){\circle*{0.5}}
\put(54.6,39.4){\circle*{0.5}}
\put(54.9,12.5){\circle*{0.5}}
\put(54.9,39.8){\circle*{0.5}}
\put(55.2,12.5){\circle*{0.5}}
\put(55.2,40.3){\circle*{0.5}}
\put(55.5,12.5){\circle*{0.5}}
\put(55.5,40.7){\circle*{0.5}}
\put(55.8,12.5){\circle*{0.5}}
\put(55.8,41.1){\circle*{0.5}}
\put(56.1,12.6){\circle*{0.5}}
\put(56.1,41.6){\circle*{0.5}}
\put(56.4,12.6){\circle*{0.5}}
\put(56.4,42.1){\circle*{0.5}}
\put(56.7,12.6){\circle*{0.5}}
\put(56.7,42.5){\circle*{0.5}}
\put(57.0,12.7){\circle*{0.5}}
\put(57.0,43.0){\circle*{0.5}}
\put(57.3,12.7){\circle*{0.5}}
\put(57.3,43.5){\circle*{0.5}}
\put(57.6,12.7){\circle*{0.5}}
\put(57.6,43.9){\circle*{0.5}}
\put(57.9,12.7){\circle*{0.5}}
\put(57.9,44.4){\circle*{0.5}}
\put(58.2,12.8){\circle*{0.5}}
\put(58.2,44.9){\circle*{0.5}}
\put(58.5,12.8){\circle*{0.5}}
\put(58.5,45.4){\circle*{0.5}}
\put(58.8,12.8){\circle*{0.5}}
\put(58.8,45.9){\circle*{0.5}}
\put(59.1,12.8){\circle*{0.5}}
\put(59.1,46.4){\circle*{0.5}}
\put(59.4,12.9){\circle*{0.5}}
\put(59.4,46.9){\circle*{0.5}}
\put(59.7,12.9){\circle*{0.5}}
\put(59.7,47.4){\circle*{0.5}}
\put(60.0,12.9){\circle*{0.5}}
\put(60.0,48.0){\circle*{0.5}}
\put(60.3,12.9){\circle*{0.5}}
\put(60.3,48.5){\circle*{0.5}}
\put(60.6,12.9){\circle*{0.5}}
\put(60.6,49.0){\circle*{0.5}}
\put(60.9,13.0){\circle*{0.5}}
\put(60.9,49.6){\circle*{0.5}}
\put(61.2,13.0){\circle*{0.5}}
\put(61.5,13.0){\circle*{0.5}}
\put(61.8,13.0){\circle*{0.5}}
\put(62.1,13.1){\circle*{0.5}}
\put(62.4,13.1){\circle*{0.5}}
\put(62.7,13.1){\circle*{0.5}}
\put(63.0,13.1){\circle*{0.5}}
\put(63.3,13.1){\circle*{0.5}}
\put(63.6,13.2){\circle*{0.5}}
\put(63.9,13.2){\circle*{0.5}}
\put(64.2,13.2){\circle*{0.5}}
\put(64.5,13.2){\circle*{0.5}}
\put(64.8,13.2){\circle*{0.5}}
\put(65.1,13.3){\circle*{0.5}}
\put(65.4,13.3){\circle*{0.5}}
\put(65.7,13.3){\circle*{0.5}}
\put(66.0,13.3){\circle*{0.5}}
\put(66.3,13.3){\circle*{0.5}}
\put(66.6,13.4){\circle*{0.5}}
\put(66.9,13.4){\circle*{0.5}}
\put(67.2,13.4){\circle*{0.5}}
\put(67.5,13.4){\circle*{0.5}}
\put(67.8,13.4){\circle*{0.5}}
\put(68.1,13.4){\circle*{0.5}}
\put(68.4,13.5){\circle*{0.5}}
\put(68.7,13.5){\circle*{0.5}}
\put(69.0,13.5){\circle*{0.5}}
\put(69.3,13.5){\circle*{0.5}}
\put(69.6,13.5){\circle*{0.5}}
\put(69.9,13.5){\circle*{0.5}}
\put(70.2,13.6){\circle*{0.5}}
\put(70.5,13.6){\circle*{0.5}}
\put(70.8,13.6){\circle*{0.5}}
\put(71.1,13.6){\circle*{0.5}}
\put(71.4,13.6){\circle*{0.5}}
\put(71.7,13.6){\circle*{0.5}}
\put(72.0,13.7){\circle*{0.5}}
\put(72.3,13.7){\circle*{0.5}}
\put(72.6,13.7){\circle*{0.5}}
\put(72.9,13.7){\circle*{0.5}}
\put(73.2,13.7){\circle*{0.5}}
\put(73.5,13.7){\circle*{0.5}}
\put(73.8,13.7){\circle*{0.5}}
\put(74.1,13.8){\circle*{0.5}}
\put(74.4,13.8){\circle*{0.5}}
\put(74.7,13.8){\circle*{0.5}}
\put(75.0,13.8){\circle*{0.5}}
\put(75.3,13.8){\circle*{0.5}}
\put(75.6,13.8){\circle*{0.5}}
\put(75.9,13.8){\circle*{0.5}}
\put(76.2,13.9){\circle*{0.5}}
\put(76.5,13.9){\circle*{0.5}}
\put(76.8,13.9){\circle*{0.5}}
\put(77.1,13.9){\circle*{0.5}}
\put(77.4,13.9){\circle*{0.5}}
\put(77.7,13.9){\circle*{0.5}}
\put(78.0,13.9){\circle*{0.5}}
\put(78.3,13.9){\circle*{0.5}}
\put(78.6,14.0){\circle*{0.5}}
\put(78.9,14.0){\circle*{0.5}}
\put(79.2,14.0){\circle*{0.5}}
\put(79.5,14.0){\circle*{0.5}}
\put(79.8,14.0){\circle*{0.5}}
\put(80.1,14.0){\circle*{0.5}}
\put(80.4,14.0){\circle*{0.5}}
\put(80.7,14.0){\circle*{0.5}}
\put(81.0,14.0){\circle*{0.5}}
\put(81.3,14.1){\circle*{0.5}}
\put(81.6,14.1){\circle*{0.5}}
\put(81.9,14.1){\circle*{0.5}}
\put(82.2,14.1){\circle*{0.5}}
\put(82.5,14.1){\circle*{0.5}}
\put(82.8,14.1){\circle*{0.5}}
\put(83.1,14.1){\circle*{0.5}}
\put(83.4,14.1){\circle*{0.5}}
\put(83.7,14.1){\circle*{0.5}}
\put(84.0,14.1){\circle*{0.5}}
\put(84.3,14.2){\circle*{0.5}}
\put(84.6,14.2){\circle*{0.5}}
\put(84.9,14.2){\circle*{0.5}}
\put(85.2,14.2){\circle*{0.5}}
\put(85.5,14.2){\circle*{0.5}}
\put(85.8,14.2){\circle*{0.5}}
\put(86.1,14.2){\circle*{0.5}}
\put(86.4,14.2){\circle*{0.5}}
\put(86.7,14.2){\circle*{0.5}}
\put(87.0,14.2){\circle*{0.5}}
\put(87.3,14.2){\circle*{0.5}}
\put(87.6,14.3){\circle*{0.5}}
\put(87.9,14.3){\circle*{0.5}}
\put(88.2,14.3){\circle*{0.5}}
\put(88.5,14.3){\circle*{0.5}}
\put(88.8,14.3){\circle*{0.5}}
\put(89.1,14.3){\circle*{0.5}}
\put(89.4,14.3){\circle*{0.5}}
\put(89.7,14.3){\circle*{0.5}}
\put(90.0,14.3){\circle*{0.5}}
\put(90.3,14.3){\circle*{0.5}}
\put(90.6,14.3){\circle*{0.5}}
\put(90.9,14.3){\circle*{0.5}}
\put(91.2,14.3){\circle*{0.5}}
\put(91.5,14.4){\circle*{0.5}}
\put(91.8,14.4){\circle*{0.5}}
\put(92.1,14.4){\circle*{0.5}}
\put(92.4,14.4){\circle*{0.5}}
\put(92.7,14.4){\circle*{0.5}}
\put(93.0,14.4){\circle*{0.5}}
\put(93.3,14.4){\circle*{0.5}}
\put(93.6,14.4){\circle*{0.5}}
\put(93.9,14.4){\circle*{0.5}}
\put(94.2,14.4){\circle*{0.5}}
\put(94.5,14.4){\circle*{0.5}}
\put(94.8,14.4){\circle*{0.5}}
\put(95.1,14.4){\circle*{0.5}}
\put(95.4,14.4){\circle*{0.5}}
\put(95.7,14.4){\circle*{0.5}}
\put(96.0,14.5){\circle*{0.5}}
\put(96.3,14.5){\circle*{0.5}}
\put(96.6,14.5){\circle*{0.5}}
\put(96.9,14.5){\circle*{0.5}}
\put(97.2,14.5){\circle*{0.5}}
\put(97.5,14.5){\circle*{0.5}}
\put(97.8,14.5){\circle*{0.5}}
\put(98.1,14.5){\circle*{0.5}}
\put(98.4,14.5){\circle*{0.5}}
\put(98.7,14.5){\circle*{0.5}}
\put(99.0,14.5){\circle*{0.5}}
\put(99.3,14.5){\circle*{0.5}}
\put(99.6,14.5){\circle*{0.5}}
\put(99.9,14.5){\circle*{0.5}}
\put(58.9,44.6){$K(t)$}
\put(99.9,16.5){$\phi(t)$}
}
\end{picture}
\caption{The scalar factor $K(t)$ and the scalar field $\phi (t)$
for the periodic potential.}
\end{figure}

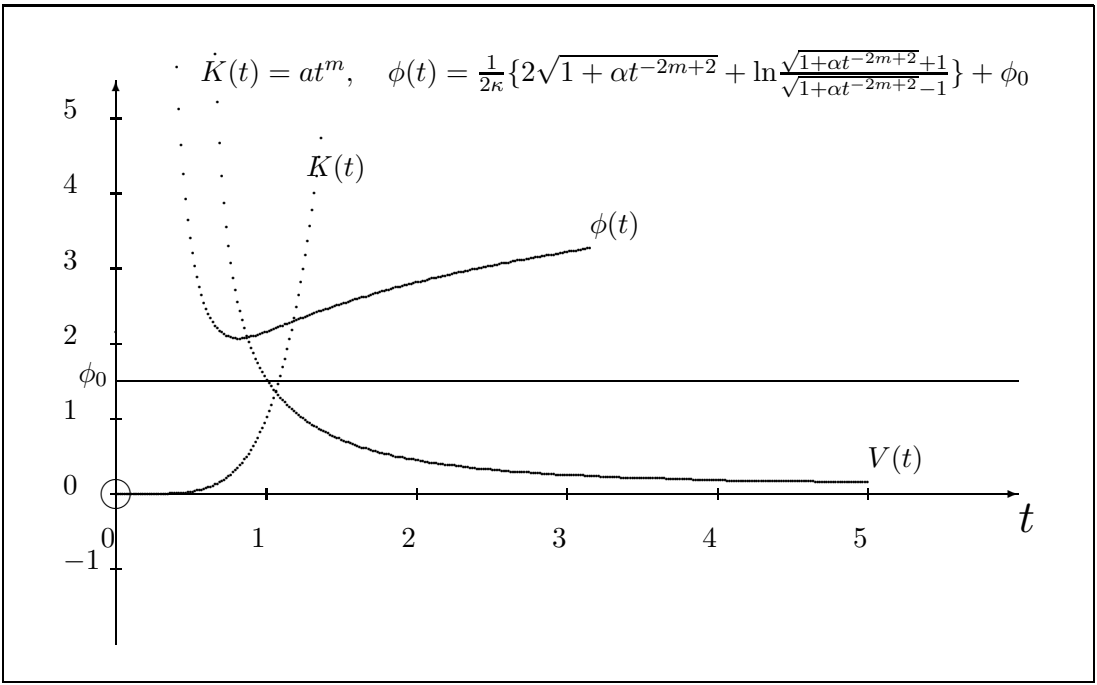
\begin{figure}  \label{Pic3}
\begin{picture}(145.0,90.0)
\put(0,0){\line(1,0){145.0}}
\put(0,90.0){\line(1,0){145.0}}
\put(0,0){\line(0,1){90.0}}
\put(145.0,0){\line(0,1){90.0}}
\put(15.0, 5.0){\vector(0,1){75.0}}
\put(15.0,25.0){\vector(1,0){120.0}}
\put(18.0,74.0){{\Large $\ $}}
\put(135.0,20.0){{\Large $t$}}
\put(25.0,80.0){{ $K(t)=at^m, \quad \phi(t)=\frac{1}{2\kappa}\{2\sqrt{1+\alpha{t}^{-2m+2}}+{\rm ln}\frac{\sqrt{1+\alpha{t}^{-2m+2}}+1}{\sqrt{1+\alpha{t}^{-2m+2}}-1}\}+\phi_0$}}
\put(15.0,25.0){
\put( 0.0,-0.7){\line(0,1){ 1.4}}
\put(-2.0,-7.0){$0$}
\put(20.0,-0.7){\line(0,1){ 1.4}}
\put(18.0,-7.0){$1$}
\put(40.0,-0.7){\line(0,1){ 1.4}}
\put(38.0,-7.0){$2$}
\put(60.0,-0.7){\line(0,1){ 1.4}}
\put(58.0,-7.0){$3$}
\put(80.0,-0.7){\line(0,1){ 1.4}}
\put(78.0,-7.0){$4$}
\put(100.0,-0.7){\line(0,1){ 1.4}}
\put(98.0,-7.0){$5$}
\put(-0.7,-10.0){\line(1,0){ 1.4}}
\put(-7.0,-10.0){$-1$}
\put(-0.7, 0.0){\line(1,0){ 1.4}}
\put(-7.0, 0.0){$0$}
\put(-0.7,10.0){\line(1,0){ 1.4}}
\put(-7.0,10.0){$1$}
\put(-0.7,20.0){\line(1,0){ 1.4}}
\put(-7.0,20.0){$2$}
\put(-0.7,30.0){\line(1,0){ 1.4}}
\put(-7.0,30.0){$3$}
\put(-0.7,40.0){\line(1,0){ 1.4}}
\put(-7.0,40.0){$4$}
\put(-0.7,50.0){\line(1,0){ 1.4}}
\put(-7.0,50.0){$5$}
\put( 0.0,15.0){\line(1,0){120.0}}\put(-5.0,15.0){$\phi_0$}\put( 0.0, 0.0){\circle*{0.5}}
\put( 0.0, 0.0){\circle{4}}
\put( 0.0,21.6){\circle*{0.5}}
\put( 0.3, 0.0){\circle*{0.5}}
\put( 0.6, 0.0){\circle*{0.5}}
\put( 0.9, 0.0){\circle*{0.5}}
\put( 1.2, 0.0){\circle*{0.5}}
\put( 1.5, 0.0){\circle*{0.5}}
\put( 1.8, 0.0){\circle*{0.5}}
\put( 2.1, 0.0){\circle*{0.5}}
\put( 2.4, 0.0){\circle*{0.5}}
\put( 2.7, 0.0){\circle*{0.5}}
\put( 3.0, 0.0){\circle*{0.5}}
\put( 3.3, 0.0){\circle*{0.5}}
\put( 3.6, 0.0){\circle*{0.5}}
\put( 3.9, 0.0){\circle*{0.5}}
\put( 4.2, 0.0){\circle*{0.5}}
\put( 4.5, 0.0){\circle*{0.5}}
\put( 4.8, 0.0){\circle*{0.5}}
\put( 5.1, 0.0){\circle*{0.5}}
\put( 5.4, 0.0){\circle*{0.5}}
\put( 5.7, 0.0){\circle*{0.5}}
\put( 6.0, 0.0){\circle*{0.5}}
\put( 6.3, 0.0){\circle*{0.5}}
\put( 6.6, 0.0){\circle*{0.5}}
\put( 6.9, 0.0){\circle*{0.5}}
\put( 7.2, 0.1){\circle*{0.5}}
\put( 7.5, 0.1){\circle*{0.5}}
\put( 7.8, 0.1){\circle*{0.5}}
\put( 8.1, 0.1){\circle*{0.5}}
\put( 8.1,56.8){\circle*{0.5}}
\put( 8.4, 0.1){\circle*{0.5}}
\put( 8.4,51.2){\circle*{0.5}}
\put( 8.7, 0.2){\circle*{0.5}}
\put( 8.7,46.5){\circle*{0.5}}
\put( 9.0, 0.2){\circle*{0.5}}
\put( 9.0,42.6){\circle*{0.5}}
\put( 9.3, 0.2){\circle*{0.5}}
\put( 9.3,39.3){\circle*{0.5}}
\put( 9.6, 0.3){\circle*{0.5}}
\put( 9.6,36.5){\circle*{0.5}}
\put( 9.9, 0.3){\circle*{0.5}}
\put( 9.9,34.1){\circle*{0.5}}
\put(10.2, 0.3){\circle*{0.5}}
\put(10.2,32.1){\circle*{0.5}}
\put(10.5, 0.4){\circle*{0.5}}
\put(10.5,30.4){\circle*{0.5}}
\put(10.8, 0.5){\circle*{0.5}}
\put(10.8,28.9){\circle*{0.5}}
\put(11.1, 0.5){\circle*{0.5}}
\put(11.1,27.6){\circle*{0.5}}
\put(11.4, 0.6){\circle*{0.5}}
\put(11.4,26.5){\circle*{0.5}}
\put(11.7, 0.7){\circle*{0.5}}
\put(11.7,25.5){\circle*{0.5}}
\put(12.0, 0.8){\circle*{0.5}}
\put(12.0,24.7){\circle*{0.5}}
\put(12.3, 0.9){\circle*{0.5}}
\put(12.3,24.0){\circle*{0.5}}
\put(12.6,   1){\circle*{0.5}}
\put(12.6,23.4){\circle*{0.5}}
\put(12.9, 1.1){\circle*{0.5}}
\put(12.9,22.9){\circle*{0.5}}
\put(13.2, 1.3){\circle*{0.5}}
\put(13.2,22.4){\circle*{0.5}}
\put(13.2,58.6){\circle*{0.5}}
\put(13.5, 1.4){\circle*{0.5}}
\put(13.5,22.1){\circle*{0.5}}
\put(13.5,52.1){\circle*{0.5}}
\put(13.8, 1.6){\circle*{0.5}}
\put(13.8,21.7){\circle*{0.5}}
\put(13.8,46.8){\circle*{0.5}}
\put(14.1, 1.7){\circle*{0.5}}
\put(14.1,21.5){\circle*{0.5}}
\put(14.1,42.4){\circle*{0.5}}
\put(14.4, 1.9){\circle*{0.5}}
\put(14.4,21.3){\circle*{0.5}}
\put(14.4,38.7){\circle*{0.5}}
\put(14.7, 2.1){\circle*{0.5}}
\put(14.7,21.1){\circle*{0.5}}
\put(14.7,35.6){\circle*{0.5}}
\put(15.0, 2.4){\circle*{0.5}}
\put(15.0,21.0){\circle*{0.5}}
\put(15.0,33.0){\circle*{0.5}}
\put(15.3, 2.6){\circle*{0.5}}
\put(15.3,20.9){\circle*{0.5}}
\put(15.3,30.7){\circle*{0.5}}
\put(15.6, 2.9){\circle*{0.5}}
\put(15.6,20.8){\circle*{0.5}}
\put(15.6,28.8){\circle*{0.5}}
\put(15.9, 3.2){\circle*{0.5}}
\put(15.9,20.8){\circle*{0.5}}
\put(15.9,27.1){\circle*{0.5}}
\put(16.2, 3.5){\circle*{0.5}}
\put(16.2,20.7){\circle*{0.5}}
\put(16.2,25.6){\circle*{0.5}}
\put(16.5, 3.8){\circle*{0.5}}
\put(16.5,20.7){\circle*{0.5}}
\put(16.5,24.3){\circle*{0.5}}
\put(16.8, 4.2){\circle*{0.5}}
\put(16.8,20.8){\circle*{0.5}}
\put(16.8,23.1){\circle*{0.5}}
\put(17.1, 4.6){\circle*{0.5}}
\put(17.1,20.8){\circle*{0.5}}
\put(17.1,22.1){\circle*{0.5}}
\put(17.4, 5.0){\circle*{0.5}}
\put(17.4,20.8){\circle*{0.5}}
\put(17.4,21.1){\circle*{0.5}}
\put(17.7, 5.4){\circle*{0.5}}
\put(17.7,20.9){\circle*{0.5}}
\put(17.7,20.2){\circle*{0.5}}
\put(18.0, 5.9){\circle*{0.5}}
\put(18.0,21.0){\circle*{0.5}}
\put(18.0,19.4){\circle*{0.5}}
\put(18.3, 6.4){\circle*{0.5}}
\put(18.3,21.0){\circle*{0.5}}
\put(18.3,18.7){\circle*{0.5}}
\put(18.6, 7.0){\circle*{0.5}}
\put(18.6,21.1){\circle*{0.5}}
\put(18.6,18.0){\circle*{0.5}}
\put(18.9, 7.5){\circle*{0.5}}
\put(18.9,21.2){\circle*{0.5}}
\put(18.9,17.4){\circle*{0.5}}
\put(19.2, 8.2){\circle*{0.5}}
\put(19.2,21.3){\circle*{0.5}}
\put(19.2,16.8){\circle*{0.5}}
\put(19.5, 8.8){\circle*{0.5}}
\put(19.5,21.4){\circle*{0.5}}
\put(19.5,16.2){\circle*{0.5}}
\put(19.8, 9.5){\circle*{0.5}}
\put(19.8,21.5){\circle*{0.5}}
\put(19.8,15.7){\circle*{0.5}}
\put(20.1,10.3){\circle*{0.5}}
\put(20.1,21.6){\circle*{0.5}}
\put(20.1,15.2){\circle*{0.5}}
\put(20.4,11.0){\circle*{0.5}}
\put(20.4,21.7){\circle*{0.5}}
\put(20.4,14.8){\circle*{0.5}}
\put(20.7,11.9){\circle*{0.5}}
\put(20.7,21.9){\circle*{0.5}}
\put(20.7,14.4){\circle*{0.5}}
\put(21.0,12.8){\circle*{0.5}}
\put(21.0,22.0){\circle*{0.5}}
\put(21.0,13.9){\circle*{0.5}}
\put(21.3,13.7){\circle*{0.5}}
\put(21.3,22.1){\circle*{0.5}}
\put(21.3,13.6){\circle*{0.5}}
\put(21.6,14.7){\circle*{0.5}}
\put(21.6,22.2){\circle*{0.5}}
\put(21.6,13.2){\circle*{0.5}}
\put(21.9,15.7){\circle*{0.5}}
\put(21.9,22.3){\circle*{0.5}}
\put(21.9,12.8){\circle*{0.5}}
\put(22.2,16.9){\circle*{0.5}}
\put(22.2,22.4){\circle*{0.5}}
\put(22.2,12.5){\circle*{0.5}}
\put(22.5,18.0){\circle*{0.5}}
\put(22.5,22.6){\circle*{0.5}}
\put(22.5,12.2){\circle*{0.5}}
\put(22.8,19.3){\circle*{0.5}}
\put(22.8,22.7){\circle*{0.5}}
\put(22.8,11.9){\circle*{0.5}}
\put(23.1,20.6){\circle*{0.5}}
\put(23.1,22.8){\circle*{0.5}}
\put(23.1,11.6){\circle*{0.5}}
\put(23.4,21.9){\circle*{0.5}}
\put(23.4,22.9){\circle*{0.5}}
\put(23.4,11.3){\circle*{0.5}}
\put(23.7,23.4){\circle*{0.5}}
\put(23.7,23.0){\circle*{0.5}}
\put(23.7,11.0){\circle*{0.5}}
\put(24.0,24.9){\circle*{0.5}}
\put(24.0,23.2){\circle*{0.5}}
\put(24.0,10.8){\circle*{0.5}}
\put(24.3,26.5){\circle*{0.5}}
\put(24.3,23.3){\circle*{0.5}}
\put(24.3,10.5){\circle*{0.5}}
\put(24.6,28.2){\circle*{0.5}}
\put(24.6,23.4){\circle*{0.5}}
\put(24.6,10.3){\circle*{0.5}}
\put(24.9,29.9){\circle*{0.5}}
\put(24.9,23.5){\circle*{0.5}}
\put(24.9,10.1){\circle*{0.5}}
\put(25.2,31.8){\circle*{0.5}}
\put(25.2,23.6){\circle*{0.5}}
\put(25.2, 9.9){\circle*{0.5}}
\put(25.5,33.7){\circle*{0.5}}
\put(25.5,23.7){\circle*{0.5}}
\put(25.5, 9.6){\circle*{0.5}}
\put(25.8,35.7){\circle*{0.5}}
\put(25.8,23.8){\circle*{0.5}}
\put(25.8, 9.4){\circle*{0.5}}
\put(26.1,37.8){\circle*{0.5}}
\put(26.1,24.0){\circle*{0.5}}
\put(26.1, 9.2){\circle*{0.5}}
\put(26.4,40.1){\circle*{0.5}}
\put(26.4,24.1){\circle*{0.5}}
\put(26.4, 9.1){\circle*{0.5}}
\put(26.7,42.4){\circle*{0.5}}
\put(26.7,24.2){\circle*{0.5}}
\put(26.7, 8.9){\circle*{0.5}}
\put(27.0,44.8){\circle*{0.5}}
\put(27.0,24.3){\circle*{0.5}}
\put(27.0, 8.7){\circle*{0.5}}
\put(27.3,47.4){\circle*{0.5}}
\put(27.3,24.4){\circle*{0.5}}
\put(27.3, 8.5){\circle*{0.5}}
\put(27.6,24.5){\circle*{0.5}}
\put(27.6, 8.4){\circle*{0.5}}
\put(27.9,24.6){\circle*{0.5}}
\put(27.9, 8.2){\circle*{0.5}}
\put(28.2,24.7){\circle*{0.5}}
\put(28.2, 8.1){\circle*{0.5}}
\put(28.5,24.8){\circle*{0.5}}
\put(28.5, 7.9){\circle*{0.5}}
\put(28.8,24.9){\circle*{0.5}}
\put(28.8, 7.8){\circle*{0.5}}
\put(29.1,25.0){\circle*{0.5}}
\put(29.1, 7.6){\circle*{0.5}}
\put(29.4,25.1){\circle*{0.5}}
\put(29.4, 7.5){\circle*{0.5}}
\put(29.7,25.2){\circle*{0.5}}
\put(29.7, 7.4){\circle*{0.5}}
\put(30.0,25.3){\circle*{0.5}}
\put(30.0, 7.2){\circle*{0.5}}
\put(30.3,25.4){\circle*{0.5}}
\put(30.3, 7.1){\circle*{0.5}}
\put(30.6,25.5){\circle*{0.5}}
\put(30.6, 7.0){\circle*{0.5}}
\put(30.9,25.6){\circle*{0.5}}
\put(30.9, 6.9){\circle*{0.5}}
\put(31.2,25.7){\circle*{0.5}}
\put(31.2, 6.8){\circle*{0.5}}
\put(31.5,25.8){\circle*{0.5}}
\put(31.5, 6.6){\circle*{0.5}}
\put(31.8,25.9){\circle*{0.5}}
\put(31.8, 6.5){\circle*{0.5}}
\put(32.1,26.0){\circle*{0.5}}
\put(32.1, 6.4){\circle*{0.5}}
\put(32.4,26.1){\circle*{0.5}}
\put(32.4, 6.3){\circle*{0.5}}
\put(32.7,26.2){\circle*{0.5}}
\put(32.7, 6.2){\circle*{0.5}}
\put(33.0,26.3){\circle*{0.5}}
\put(33.0, 6.1){\circle*{0.5}}
\put(33.3,26.3){\circle*{0.5}}
\put(33.3, 6.1){\circle*{0.5}}
\put(33.6,26.4){\circle*{0.5}}
\put(33.6, 6.0){\circle*{0.5}}
\put(33.9,26.5){\circle*{0.5}}
\put(33.9, 5.9){\circle*{0.5}}
\put(34.2,26.6){\circle*{0.5}}
\put(34.2, 5.8){\circle*{0.5}}
\put(34.5,26.7){\circle*{0.5}}
\put(34.5, 5.7){\circle*{0.5}}
\put(34.8,26.8){\circle*{0.5}}
\put(34.8, 5.6){\circle*{0.5}}
\put(35.1,26.9){\circle*{0.5}}
\put(35.1, 5.5){\circle*{0.5}}
\put(35.4,27.0){\circle*{0.5}}
\put(35.4, 5.5){\circle*{0.5}}
\put(35.7,27.0){\circle*{0.5}}
\put(35.7, 5.4){\circle*{0.5}}
\put(36.0,27.1){\circle*{0.5}}
\put(36.0, 5.3){\circle*{0.5}}
\put(36.3,27.2){\circle*{0.5}}
\put(36.3, 5.3){\circle*{0.5}}
\put(36.6,27.3){\circle*{0.5}}
\put(36.6, 5.2){\circle*{0.5}}
\put(36.9,27.4){\circle*{0.5}}
\put(36.9, 5.1){\circle*{0.5}}
\put(37.2,27.5){\circle*{0.5}}
\put(37.2, 5.0){\circle*{0.5}}
\put(37.5,27.5){\circle*{0.5}}
\put(37.5, 5.0){\circle*{0.5}}
\put(37.8,27.6){\circle*{0.5}}
\put(37.8, 4.9){\circle*{0.5}}
\put(38.1,27.7){\circle*{0.5}}
\put(38.1, 4.9){\circle*{0.5}}
\put(38.4,27.8){\circle*{0.5}}
\put(38.4, 4.8){\circle*{0.5}}
\put(38.7,27.8){\circle*{0.5}}
\put(38.7, 4.7){\circle*{0.5}}
\put(39.0,27.9){\circle*{0.5}}
\put(39.0, 4.7){\circle*{0.5}}
\put(39.3,28.0){\circle*{0.5}}
\put(39.3, 4.6){\circle*{0.5}}
\put(39.6,28.1){\circle*{0.5}}
\put(39.6, 4.6){\circle*{0.5}}
\put(39.9,28.2){\circle*{0.5}}
\put(39.9, 4.5){\circle*{0.5}}
\put(40.2,28.2){\circle*{0.5}}
\put(40.2, 4.5){\circle*{0.5}}
\put(40.5,28.3){\circle*{0.5}}
\put(40.5, 4.4){\circle*{0.5}}
\put(40.8,28.4){\circle*{0.5}}
\put(40.8, 4.4){\circle*{0.5}}
\put(41.1,28.4){\circle*{0.5}}
\put(41.1, 4.3){\circle*{0.5}}
\put(41.4,28.5){\circle*{0.5}}
\put(41.4, 4.3){\circle*{0.5}}
\put(41.7,28.6){\circle*{0.5}}
\put(41.7, 4.2){\circle*{0.5}}
\put(42.0,28.7){\circle*{0.5}}
\put(42.0, 4.2){\circle*{0.5}}
\put(42.3,28.7){\circle*{0.5}}
\put(42.3, 4.1){\circle*{0.5}}
\put(42.6,28.8){\circle*{0.5}}
\put(42.6, 4.1){\circle*{0.5}}
\put(42.9,28.9){\circle*{0.5}}
\put(42.9, 4.0){\circle*{0.5}}
\put(43.2,28.9){\circle*{0.5}}
\put(43.2, 4.0){\circle*{0.5}}
\put(43.5,29.0){\circle*{0.5}}
\put(43.5, 4.0){\circle*{0.5}}
\put(43.8,29.1){\circle*{0.5}}
\put(43.8, 3.9){\circle*{0.5}}
\put(44.1,29.2){\circle*{0.5}}
\put(44.1, 3.9){\circle*{0.5}}
\put(44.4,29.2){\circle*{0.5}}
\put(44.4, 3.8){\circle*{0.5}}
\put(44.7,29.3){\circle*{0.5}}
\put(44.7, 3.8){\circle*{0.5}}
\put(45.0,29.4){\circle*{0.5}}
\put(45.0, 3.8){\circle*{0.5}}
\put(45.3,29.4){\circle*{0.5}}
\put(45.3, 3.7){\circle*{0.5}}
\put(45.6,29.5){\circle*{0.5}}
\put(45.6, 3.7){\circle*{0.5}}
\put(45.9,29.6){\circle*{0.5}}
\put(45.9, 3.7){\circle*{0.5}}
\put(46.2,29.6){\circle*{0.5}}
\put(46.2, 3.6){\circle*{0.5}}
\put(46.5,29.7){\circle*{0.5}}
\put(46.5, 3.6){\circle*{0.5}}
\put(46.8,29.7){\circle*{0.5}}
\put(46.8, 3.6){\circle*{0.5}}
\put(47.1,29.8){\circle*{0.5}}
\put(47.1, 3.5){\circle*{0.5}}
\put(47.4,29.9){\circle*{0.5}}
\put(47.4, 3.5){\circle*{0.5}}
\put(47.7,29.9){\circle*{0.5}}
\put(47.7, 3.5){\circle*{0.5}}
\put(48.0,30.0){\circle*{0.5}}
\put(48.0, 3.4){\circle*{0.5}}
\put(48.3,30.1){\circle*{0.5}}
\put(48.3, 3.4){\circle*{0.5}}
\put(48.6,30.1){\circle*{0.5}}
\put(48.6, 3.4){\circle*{0.5}}
\put(48.9,30.2){\circle*{0.5}}
\put(48.9, 3.3){\circle*{0.5}}
\put(49.2,30.2){\circle*{0.5}}
\put(49.2, 3.3){\circle*{0.5}}
\put(49.5,30.3){\circle*{0.5}}
\put(49.5, 3.3){\circle*{0.5}}
\put(49.8,30.4){\circle*{0.5}}
\put(49.8, 3.3){\circle*{0.5}}
\put(50.1,30.4){\circle*{0.5}}
\put(50.1, 3.2){\circle*{0.5}}
\put(50.4,30.5){\circle*{0.5}}
\put(50.4, 3.2){\circle*{0.5}}
\put(50.7,30.5){\circle*{0.5}}
\put(50.7, 3.2){\circle*{0.5}}
\put(51.0,30.6){\circle*{0.5}}
\put(51.0, 3.2){\circle*{0.5}}
\put(51.3,30.7){\circle*{0.5}}
\put(51.3, 3.1){\circle*{0.5}}
\put(51.6,30.7){\circle*{0.5}}
\put(51.6, 3.1){\circle*{0.5}}
\put(51.9,30.8){\circle*{0.5}}
\put(51.9, 3.1){\circle*{0.5}}
\put(52.2,30.8){\circle*{0.5}}
\put(52.2, 3.1){\circle*{0.5}}
\put(52.5,30.9){\circle*{0.5}}
\put(52.5, 3.0){\circle*{0.5}}
\put(52.8,31.0){\circle*{0.5}}
\put(52.8, 3.0){\circle*{0.5}}
\put(53.1,31.0){\circle*{0.5}}
\put(53.1, 3.0){\circle*{0.5}}
\put(53.4,31.1){\circle*{0.5}}
\put(53.4, 3.0){\circle*{0.5}}
\put(53.7,31.1){\circle*{0.5}}
\put(53.7, 2.9){\circle*{0.5}}
\put(54.0,31.2){\circle*{0.5}}
\put(54.0, 2.9){\circle*{0.5}}
\put(54.3,31.2){\circle*{0.5}}
\put(54.3, 2.9){\circle*{0.5}}
\put(54.6,31.3){\circle*{0.5}}
\put(54.6, 2.9){\circle*{0.5}}
\put(54.9,31.3){\circle*{0.5}}
\put(54.9, 2.9){\circle*{0.5}}
\put(55.2,31.4){\circle*{0.5}}
\put(55.2, 2.8){\circle*{0.5}}
\put(55.5,31.4){\circle*{0.5}}
\put(55.5, 2.8){\circle*{0.5}}
\put(55.8,31.5){\circle*{0.5}}
\put(55.8, 2.8){\circle*{0.5}}
\put(56.1,31.6){\circle*{0.5}}
\put(56.1, 2.8){\circle*{0.5}}
\put(56.4,31.6){\circle*{0.5}}
\put(56.4, 2.8){\circle*{0.5}}
\put(56.7,31.7){\circle*{0.5}}
\put(56.7, 2.7){\circle*{0.5}}
\put(57.0,31.7){\circle*{0.5}}
\put(57.0, 2.7){\circle*{0.5}}
\put(57.3,31.8){\circle*{0.5}}
\put(57.3, 2.7){\circle*{0.5}}
\put(57.6,31.8){\circle*{0.5}}
\put(57.6, 2.7){\circle*{0.5}}
\put(57.9,31.9){\circle*{0.5}}
\put(57.9, 2.7){\circle*{0.5}}
\put(58.2,31.9){\circle*{0.5}}
\put(58.2, 2.7){\circle*{0.5}}
\put(58.5,32.0){\circle*{0.5}}
\put(58.5, 2.6){\circle*{0.5}}
\put(58.8,32.0){\circle*{0.5}}
\put(58.8, 2.6){\circle*{0.5}}
\put(59.1,32.1){\circle*{0.5}}
\put(59.1, 2.6){\circle*{0.5}}
\put(59.4,32.1){\circle*{0.5}}
\put(59.4, 2.6){\circle*{0.5}}
\put(59.7,32.2){\circle*{0.5}}
\put(59.7, 2.6){\circle*{0.5}}
\put(60.0,32.2){\circle*{0.5}}
\put(60.0, 2.6){\circle*{0.5}}
\put(60.3,32.3){\circle*{0.5}}
\put(60.3, 2.5){\circle*{0.5}}
\put(60.6,32.3){\circle*{0.5}}
\put(60.6, 2.5){\circle*{0.5}}
\put(60.9,32.4){\circle*{0.5}}
\put(60.9, 2.5){\circle*{0.5}}
\put(61.2,32.4){\circle*{0.5}}
\put(61.2, 2.5){\circle*{0.5}}
\put(61.5,32.5){\circle*{0.5}}
\put(61.5, 2.5){\circle*{0.5}}
\put(61.8,32.5){\circle*{0.5}}
\put(61.8, 2.5){\circle*{0.5}}
\put(62.1,32.6){\circle*{0.5}}
\put(62.1, 2.5){\circle*{0.5}}
\put(62.4,32.6){\circle*{0.5}}
\put(62.4, 2.4){\circle*{0.5}}
\put(62.7,32.7){\circle*{0.5}}
\put(62.7, 2.4){\circle*{0.5}}
\put(63.0,32.7){\circle*{0.5}}
\put(63.0, 2.4){\circle*{0.5}}
\put(63.3, 2.4){\circle*{0.5}}
\put(63.6, 2.4){\circle*{0.5}}
\put(63.9, 2.4){\circle*{0.5}}
\put(64.2, 2.4){\circle*{0.5}}
\put(64.5, 2.3){\circle*{0.5}}
\put(64.8, 2.3){\circle*{0.5}}
\put(65.1, 2.3){\circle*{0.5}}
\put(65.4, 2.3){\circle*{0.5}}
\put(65.7, 2.3){\circle*{0.5}}
\put(66.0, 2.3){\circle*{0.5}}
\put(66.3, 2.3){\circle*{0.5}}
\put(66.6, 2.3){\circle*{0.5}}
\put(66.9, 2.3){\circle*{0.5}}
\put(67.2, 2.2){\circle*{0.5}}
\put(67.5, 2.2){\circle*{0.5}}
\put(67.8, 2.2){\circle*{0.5}}
\put(68.1, 2.2){\circle*{0.5}}
\put(68.4, 2.2){\circle*{0.5}}
\put(68.7, 2.2){\circle*{0.5}}
\put(69.0, 2.2){\circle*{0.5}}
\put(69.3, 2.2){\circle*{0.5}}
\put(69.6, 2.2){\circle*{0.5}}
\put(69.9, 2.1){\circle*{0.5}}
\put(70.2, 2.1){\circle*{0.5}}
\put(70.5, 2.1){\circle*{0.5}}
\put(70.8, 2.1){\circle*{0.5}}
\put(71.1, 2.1){\circle*{0.5}}
\put(71.4, 2.1){\circle*{0.5}}
\put(71.7, 2.1){\circle*{0.5}}
\put(72.0, 2.1){\circle*{0.5}}
\put(72.3, 2.1){\circle*{0.5}}
\put(72.6, 2.1){\circle*{0.5}}
\put(72.9, 2.1){\circle*{0.5}}
\put(73.2, 2.0){\circle*{0.5}}
\put(73.5, 2.0){\circle*{0.5}}
\put(73.8, 2.0){\circle*{0.5}}
\put(74.1, 2.0){\circle*{0.5}}
\put(74.4, 2.0){\circle*{0.5}}
\put(74.7, 2.0){\circle*{0.5}}
\put(75.0, 2.0){\circle*{0.5}}
\put(75.3, 2.0){\circle*{0.5}}
\put(75.6, 2.0){\circle*{0.5}}
\put(75.9, 2.0){\circle*{0.5}}
\put(76.2, 2.0){\circle*{0.5}}
\put(76.5, 2.0){\circle*{0.5}}
\put(76.8, 1.9){\circle*{0.5}}
\put(77.1, 1.9){\circle*{0.5}}
\put(77.4, 1.9){\circle*{0.5}}
\put(77.7, 1.9){\circle*{0.5}}
\put(78.0, 1.9){\circle*{0.5}}
\put(78.3, 1.9){\circle*{0.5}}
\put(78.6, 1.9){\circle*{0.5}}
\put(78.9, 1.9){\circle*{0.5}}
\put(79.2, 1.9){\circle*{0.5}}
\put(79.5, 1.9){\circle*{0.5}}
\put(79.8, 1.9){\circle*{0.5}}
\put(80.1, 1.9){\circle*{0.5}}
\put(80.4, 1.9){\circle*{0.5}}
\put(80.7, 1.9){\circle*{0.5}}
\put(81.0, 1.9){\circle*{0.5}}
\put(81.3, 1.8){\circle*{0.5}}
\put(81.6, 1.8){\circle*{0.5}}
\put(81.9, 1.8){\circle*{0.5}}
\put(82.2, 1.8){\circle*{0.5}}
\put(82.5, 1.8){\circle*{0.5}}
\put(82.8, 1.8){\circle*{0.5}}
\put(83.1, 1.8){\circle*{0.5}}
\put(83.4, 1.8){\circle*{0.5}}
\put(83.7, 1.8){\circle*{0.5}}
\put(84.0, 1.8){\circle*{0.5}}
\put(84.3, 1.8){\circle*{0.5}}
\put(84.6, 1.8){\circle*{0.5}}
\put(84.9, 1.8){\circle*{0.5}}
\put(85.2, 1.8){\circle*{0.5}}
\put(85.5, 1.8){\circle*{0.5}}
\put(85.8, 1.8){\circle*{0.5}}
\put(86.1, 1.8){\circle*{0.5}}
\put(86.4, 1.8){\circle*{0.5}}
\put(86.7, 1.7){\circle*{0.5}}
\put(87.0, 1.7){\circle*{0.5}}
\put(87.3, 1.7){\circle*{0.5}}
\put(87.6, 1.7){\circle*{0.5}}
\put(87.9, 1.7){\circle*{0.5}}
\put(88.2, 1.7){\circle*{0.5}}
\put(88.5, 1.7){\circle*{0.5}}
\put(88.8, 1.7){\circle*{0.5}}
\put(89.1, 1.7){\circle*{0.5}}
\put(89.4, 1.7){\circle*{0.5}}
\put(89.7, 1.7){\circle*{0.5}}
\put(90.0, 1.7){\circle*{0.5}}
\put(90.3, 1.7){\circle*{0.5}}
\put(90.6, 1.7){\circle*{0.5}}
\put(90.9, 1.7){\circle*{0.5}}
\put(91.2, 1.7){\circle*{0.5}}
\put(91.5, 1.7){\circle*{0.5}}
\put(91.8, 1.7){\circle*{0.5}}
\put(92.1, 1.7){\circle*{0.5}}
\put(92.4, 1.7){\circle*{0.5}}
\put(92.7, 1.7){\circle*{0.5}}
\put(93.0, 1.6){\circle*{0.5}}
\put(93.3, 1.6){\circle*{0.5}}
\put(93.6, 1.6){\circle*{0.5}}
\put(93.9, 1.6){\circle*{0.5}}
\put(94.2, 1.6){\circle*{0.5}}
\put(94.5, 1.6){\circle*{0.5}}
\put(94.8, 1.6){\circle*{0.5}}
\put(95.1, 1.6){\circle*{0.5}}
\put(95.4, 1.6){\circle*{0.5}}
\put(95.7, 1.6){\circle*{0.5}}
\put(96.0, 1.6){\circle*{0.5}}
\put(96.3, 1.6){\circle*{0.5}}
\put(96.6, 1.6){\circle*{0.5}}
\put(96.9, 1.6){\circle*{0.5}}
\put(97.2, 1.6){\circle*{0.5}}
\put(97.5, 1.6){\circle*{0.5}}
\put(97.8, 1.6){\circle*{0.5}}
\put(98.1, 1.6){\circle*{0.5}}
\put(98.4, 1.6){\circle*{0.5}}
\put(98.7, 1.6){\circle*{0.5}}
\put(99.0, 1.6){\circle*{0.5}}
\put(99.3, 1.6){\circle*{0.5}}
\put(99.6, 1.6){\circle*{0.5}}
\put(99.9, 1.6){\circle*{0.5}}
\put(25.3,42.4){$K(t)$}
\put(63.0,34.7){$\phi(t)$}
\put(99.9, 3.6){$V(t)$}
}
\end{picture}
\caption{Power law inflation with m=5}
\end{figure}

\end{document}